\documentclass[12pt]{article}
\usepackage{amsmath,amssymb,amsthm,amsxtra,bbm,overpic,bm,epsfig,subfigure}
\usepackage[colorlinks]{hyperref}
\usepackage{mathrsfs}
\usepackage{enumitem}
\usepackage{graphicx}
\usepackage{color}
\usepackage{comment}
\usepackage{epstopdf}
\usepackage{float}
\usepackage{cite}
\usepackage{upgreek,overpic,setspace,multirow,textcomp}
\numberwithin{equation}{section}
\allowdisplaybreaks[2]

\usepackage{slashed,stmaryrd}

\def\thefootnote{\fnsymbol{footnote}}

\usepackage{slashed,stmaryrd}

\usepackage{lscape}%
\usepackage{array}
\usepackage{booktabs}%

\begin{document}
	
	\vspace{0.2cm}
	
	\begin{center}
		{\Large\bf The Mikheyev-Smirnov-Wolfenstein Matter Potential at the One-loop Level in the Standard Model}
	\end{center}
	
	\vspace{0.2cm}
	
	\begin{center}
		{\bf Jihong Huang}~\footnote{E-mail: huangjh@ihep.ac.cn},
		\quad
		{\bf Shun Zhou}~\footnote{E-mail: zhoush@ihep.ac.cn (corresponding author)}
		\\
		\vspace{0.2cm}
		{\small
			Institute of High Energy Physics, Chinese Academy of Sciences, Beijing 100049, China\\
			School of Physical Sciences, University of Chinese Academy of Sciences, Beijing 100049, China}
	\end{center}

	\vspace{0.5cm}
	
	\begin{abstract}
		When neutrinos are propagating in ordinary matter, their coherent forward scattering off background particles results in the so-called Mikheyev-Smirnov-Wolfenstein (MSW) matter potential, which plays an important role in neutrino flavor conversions. In this paper, we present a complete one-loop calculation of the MSW matter potential in the Standard Model (SM). First, we carry out the one-loop renormalization of the SM in the on-shell scheme, where the electromagnetic fine-structure constant $\alpha$, the weak gauge-boson masses $m^{}_W$ and $m^{}_Z$, the Higgs-boson mass $m^{}_h$ and the fermion masses $m^{}_f$ are chosen as input parameters. Then, the finite corrections to the scattering amplitudes of neutrinos with the electrons and quarks are calculated, and the one-loop MSW matter potentials are derived. Adopting the latest values of all physical parameters, we find that the relative size of one-loop correction to the charged-current matter potential of electron-type neutrinos or antineutrinos turns out to be $6\%$, whereas that to the neutral-current matter potential of all-flavor neutrinos or antineutrinos can be as large as $8\%$. The calculations are also performed in the $\overline{\rm MS}$ scheme and compared with previous results in the literature.
  
	\end{abstract}
	
	
	
	\def\thefootnote{\arabic{footnote}}
	\setcounter{footnote}{0}
	
	\newpage
	
	\section{Introduction}
	
	In the past quarter of a century, neutrino oscillation experiments have provided us with robust evidence that neutrinos are massive and leptonic flavor mixing is significant~\cite{ParticleDataGroup:2022pth,Xing:2020ijf}. For the neutrinos propagating in matter, the coherent forward scattering of neutrinos off the background particles leads to the Mikheyev-Smirnov-Wolfenstein (MSW) matter potential and could modify neutrino flavor conversions in a remarkable way~\cite{Wolfenstein:1977ue, Wolfenstein:1979ni, Mikheyev:1985zog, Mikheev:1986wj}. To be explicit, at the tree level in the Standard Model (SM), the effective Hamiltonian for neutrino oscillations in matter receives extra potential terms, i.e., ${\cal V}_e^{} = {\cal V}_{\rm CC}^{} + {\cal V}_{\rm NC}^{}$ for electron neutrinos and ${\cal V}_{\mu}^{} = {\cal V}_{\tau}^{} = {\cal V}_{\rm NC}^{}$ for muon and tau neutrinos, where the charged-current (CC) and the neutral-current (NC) contributions are given by
	\begin{eqnarray} \label{eq:Vtree}
		{\cal V}_{\rm CC}^{} = \sqrt{2} G_{\mu}^{} N_e^{}\;, \qquad
		{\cal V}_{\rm NC}^{} = -\frac{G_{\mu}^{}}{\sqrt{2}} \left[\left(1 - 4 \sin^2\theta_{\rm w}^{}\right) \left(N_e^{} - N_p^{}\right) + N_n^{}\right]\;. 
	\end{eqnarray}
	In Eq.~(\ref{eq:Vtree}), $G_{\mu}^{}$ is the Fermi constant determined from the muon lifetime, $N_e^{}, N_p^{}$ and $N_n^{}$ are respectively the net number densities of electrons, protons and neutrons, and $\theta_{\rm w}^{}$ is the weak mixing angle. For antineutrinos, the MSW matter potentials ${\cal V}^{}_\alpha$ (for $\alpha = e, \mu, \tau$) change accordingly to opposite signs. As the NC potential ${\cal V}^{}_{\rm NC}$ is universal for three neutrino flavors, only the CC potential ${\cal V}^{}_{\rm CC}$ for electron (anti)neutrinos is relevant for neutrino flavor conversions in matter.
	
	At the one-loop level in the SM, it has been known for a long time that the NC potentials $\widehat{\cal V}^{\alpha}_{\rm NC}$ become dependent on the charged-lepton masses $m^{}_\alpha$ (for $\alpha = e, \mu, \tau$). Given the strong hierarchy of charged-lepton masses $m_e^{} \ll m_\mu^{} \ll m_\tau^{} $ and $N_n^{} = N_p^{} = N_e^{}$ for ordinary matter, one can estimate the ratio of the flavor-dependent part of one-loop NC potential to the tree-level CC potential as below~\cite{Botella:1986wy,Mirizzi:2009td}
	\begin{eqnarray} \label{eq:ratioV}
		\epsilon^{}_{\mu\tau} \equiv \frac{\widehat{\cal V}^\tau_{\rm NC} - \widehat{\cal V}^\mu_{\rm NC}}{{\cal V}_{\rm CC}} \approx - \frac{3\alpha}{2\pi \sin^2\theta_{\rm w}^{}} \frac{m_\tau^2}{m_W^{2}} \left[\ln\left(\frac{m_\tau^2}{m_W^2} \right) + \frac{5}{6}\right]\;,
	\end{eqnarray}
	where $\alpha \equiv e^2/(4\pi)$ denotes the electromagnetic fine-structure constant. With the input values of $\alpha = 1/137$, $m^{}_W = 80.377~{\rm GeV}$, $m^{}_Z = 91.1876~{\rm GeV}$ and $m^{}_\tau = 1.777~{\rm GeV}$, one has $\sin^2\theta^{}_{\rm w} = 1 - m^2_W/m^2_Z \approx 0.223$ and thus finds the ratio in Eq.~(\ref{eq:ratioV}) to be $\epsilon^{}_{\mu\tau} \approx 5.19\times 10^{-5}$. Although such a correction is extremely small, it causes the difference between the matter potential of $\nu^{}_\mu$ and that of $\nu^{}_\tau$, which affects greatly the flavor conversions of supernova neutrinos in the dense-matter environment~\cite{Dutta:1999ir,Dighe:1999bi}. Further discussions about the impact of $\epsilon^{}_{\mu\tau}$ on neutrino oscillations can be found in Refs.~\cite{Zhu:2020wuy,Xing:2022efm}.
	
	In the calculation of $\epsilon^{}_{\mu\tau}$, however, the previous works~\cite{Botella:1986wy,Mirizzi:2009td} concentrate on the flavor-dependent radiative corrections, e.g., $\widehat{\cal V}^\tau_{\rm NC} - \widehat{\cal V}^\mu_{\rm NC}$, instead of the one-loop NC potentials $\widehat{\cal V}^\alpha_{\rm NC}$ themselves (for $\alpha = e, \mu, \tau$). Moreover, the one-loop radiative corrections to the CC potential in the on-shell scheme have not been studied thus far.\footnote{The radiative corrections in the $\overline{\rm MS}$ scheme have been evaluated in Ref.~\cite{Hill:2019xqk} in the low-energy effective theory. The authors are grateful to Dr. Oleksandr Tomalak for bringing this relevant work to our attention.} Therefore, it is interesting to calculate neutrino matter potentials in the SM at the one-loop level, including the NC potential $\widehat{\cal V}^\alpha_{\rm NC}$ for three-flavor neutrinos and the CC potential $\widehat{\cal V}^{}_{\rm CC}$ for the electron neutrino. The motivation for such a calculation is two-fold. First, the flavor-independent part of the one-loop NC potential $\widehat{\cal V}^\alpha_{\rm NC}$ is irrelevant for flavor oscillations of three active neutrinos, but may be important for active-sterile neutrino oscillations, particularly in the supernova environment~\cite{Tamborra:2011is,Wu:2013gxa}. Second, the future long-baseline accelerator neutrino oscillation experiments, such as DUNE~\cite{DUNE:2020ypp} and T2HK~\cite{Hyper-Kamiokande:2022smq}, will be able to determine neutrino mass ordering and probe leptonic CP violation, and they are already sensitive enough to the Earth matter effects. Obviously, the precise calculation of $\widehat{\cal V}^{}_{\rm CC}$ at the one-loop level is necessary to achieve high-precision measurements of the neutrino mass ordering and the CP-violating phase. 
	
	In this work, we carry out a complete one-loop calculation of the MSW potentials in the SM. More explicitly, after performing one-loop renormalization of the SM in the on-shell scheme~\cite{Aoki:1982ed,Bohm:1986rj,Hollik:1988ii,Denner:1991kt}, we compute the scattering amplitudes for $\nu_\alpha^{} + f \to \nu_\alpha^{} + f$ at one loop, where $f = u, d, e$ are the SM fermions in ordinary matter. For the electron neutrino $\nu_e^{}$, both CC and NC interactions must be taken into account, while only the latter is considered for $\nu_{\mu,\tau}^{}$. For both NC and CC interactions, since the distributions of background particles are assumed to be homogeneous and isotropic, only the vector-type couplings $c_{\rm V, NC}^{f}$ and $c^{f}_{\rm V, CC}$ are directly involved in matter potentials. After obtaining finite scattering amplitudes, we extract the matter potentials by comparing the obtained amplitudes and those generated by the effective weak Hamiltonian of neutrino interactions in the forward limit. After inputting the latest values of all physical parameters, we find that the one-loop correction to the NC potential is about $8\%$, while that to the CC potential is about $6\%$. In the future long-baseline accelerator neutrino oscillation experiments, e.g., DUNE and T2HK, it is promising to probe the one-loop correction to the CC potential. For comparison, we also calculate the one-loop corrections in the $\overline{\rm MS}$ scheme with running parameters as inputs. Our results of the vector-type couplings agree perfectly with those in the previous work~\cite{Hill:2019xqk}.
	
	The remaining part of this paper is organized as follows. In Sec.~\ref{sec:stra}, we outline the basic strategy for one-loop calculations of the MSW matter potentials in the SM, and explain the notations and the on-shell scheme of the one-loop renormalization implemented in our calculations. The analytical results for the one-loop NC and CC potentials are presented in Sec.~\ref{sec:VNC} and Sec.~\ref{sec:VCC}, respectively. Then, in Sec.~\ref{sec:num}, we specify the input parameters and evaluate the one-loop corrections. The calculations of the vector-type couplings at the one-loop level in the $\overline{\rm MS}$ scheme are given in Sec.~\ref{sec:msbar}. We summarize our main results in Sec.~\ref{sec:sum}. For completeness, the renormalization of the SM and some details of our calculations are given in Appendix~\ref{app:A}.
	
	\section{Strategy for One-loop Calculations}
	\label{sec:stra}
	
	In this section, we explain how to calculate the one-loop potentials in the SM. For the low-energy neutrinos propagating in ordinary matter, the coherent forward scattering with background particles modifies their dispersion relations and its impact on neutrino flavor conversions can be described by the effective potentials at the amplitude level. The ordinary matter is composed of protons, neutrons and electrons, so the NC interactions contribute to the matter potentials for all-flavor neutrinos, whereas the CC interaction is relevant only for the electron neutrinos.
	
	\subsection{Effective Hamiltonians and Matter Potentials}
	
	The amplitudes for relevant two-body scattering processes $\nu^{}_\alpha + f \to \nu^{}_\alpha + f$, with $\alpha = e, \mu, \tau$ and $f = u, d, e$, can be divided into the NC and CC parts. For the NC part, we can directly read it off from the low-energy effective Hamiltonian
	\begin{eqnarray}\label{eq:Heff}
		{\cal H}_{\rm eff}^{\rm NC} (x) = \frac{G_{\mu}^{}}{\sqrt{2}} \left[\overline{\nu_\alpha^{} (x)} \gamma^\mu \left(1-\gamma^5\right) \nu_\alpha^{} (x)\right] \left[\overline{f(x)} \gamma_\mu \left(c_{\rm V, NC}^f - c_{\rm A, NC}^f \gamma^5 \right) f (x)\right]\;,
	\end{eqnarray}
	where $c^f_{\rm V, NC}$ and $c^f_{\rm A, NC}$ refer respectively to the vector-type and axial-vector-type couplings for the NC interaction. At the tree level, these couplings in the SM have been collected in Table~\ref{tab:cv_ca}.
	\begin{table}[!t]
		\centering
		\begin{spacing}{2.15}
			\begin{tabular}{cccc}
				\hline\hline 
				& $f=u$ & $f=d$ & $f=e$ \\  \hline 
				$c_{\rm V, NC}^f$ & $\displaystyle \frac{1}{2}-\frac{4}{3}s^2$    & $\displaystyle -\frac{1}{2}+\frac{2}{3}s^2$    & $\displaystyle -\frac{1}{2}+2s^2$   \\ 
				$c_{\rm A, NC}^f$ & $\displaystyle \frac{1}{2}$    &  $\displaystyle -\frac{1}{2}$   & $\displaystyle -\frac{1}{2}$  \\ \hline\hline
			\end{tabular}
		\end{spacing}
		\caption{The vector-type and axial-vector-type couplings for the NC interaction of neutrinos in the SM, where $s \equiv 
			\sin \theta^{}_{\rm w}$ and $f = u, d, e$.}
		\label{tab:cv_ca}
	\end{table}
	
	Assuming the distribution of background fermions to be homogeneous and isotropic, one can average the effective Hamiltonian over all possible states of background fermions and then obtain the effective potential for the SM left-handed neutrinos~\cite{Giunti:2007ry,Xing:2011zza}
	\begin{eqnarray}\label{eq:Vl}
		{\cal V}_{\rm NC}^{} = \sqrt{2} G_{\mu}^{} N_f^{} c_{\rm V, NC}^f\;,
	\end{eqnarray}
	where $N^{}_f$ is the net number density of the background fermion $f$ and only the vector-type coupling $c^f_{\rm V, NC}$ is involved. Notice that the NC potential is independent of neutrino flavors at the tree level.
	
	For electron neutrinos, the CC part of the two-body scattering amplitude can be derived from the effective Hamiltonian
	\begin{eqnarray}\label{eq:Heff}
		{\cal H}_{\rm eff}^{\rm CC} (x) = \frac{G_{\mu}^{}}{\sqrt{2}} \left[\overline{\nu_e^{} (x)} \gamma^\mu \left(1 - \gamma^5 \right) \nu_e^{} (x)\right] \left[\overline{e(x)} \gamma_\mu \left(c^e_{\rm V, CC} -c^e_{\rm A, CC} \gamma^5\right) e (x)\right]\;,
	\end{eqnarray}
	where the Fierz transformation has been performed and $c
	^e_{\rm V, CC} = c^e_{\rm A, CC} = 1$ in the SM. In a similar way to the derivation of the NC potential, one can easily get the CC potential of electron neutrinos
	\begin{eqnarray}\label{eq:Ve}
		{\cal V}_{\rm CC}^{} = \sqrt{2} G_{\mu}^{} N_e^{} c^e_{\rm V, CC}\;.
	\end{eqnarray}
	Therefore, the total matter potential for electron neutrinos is ${\cal V}^{}_e = {\cal V}^{}_{\rm CC} + {\cal V}^{}_{\rm NC}$, while those for muon and tau neutrinos are ${\cal V}^{}_\mu = {\cal V}^{}_\tau = {\cal V}^{}_{\rm NC}$. For ordinary matter composed of protons, neutrons and electrons, together with the vector-type couplings in Table~\ref{tab:cv_ca}, one can simply use $N^{}_u = 2N^{}_p + N^{}_n$ and $N^{}_d = 2 N^{}_n + N^{}_p$ and the condition of charge neutrality $N^{}_p = N^{}_e$ to reproduce the results of ${\cal V}^{}_{\rm CC}$ and ${\cal V}^{}_{\rm NC}$ in Eq.~(\ref{eq:Vtree}).
	
	From the above derivations of the tree-level matter potentials, it is evident that one should calculate the renormalized scattering amplitude of $\nu^{}_\alpha + f \to \nu^{}_\alpha + f$ at the one-loop level and then find out the effective Hamiltonian corresponding to the loop-corrected amplitude. Starting with the loop-level effective Hamiltonian, we can extract the coefficient for the vector-type interactions involving the background particles. More explicitly, for the NC part, we identity the correction to the vector-type coupling  $c_{\rm V,NC}^f$, which will be denoted as $\Delta c_{\rm V,NC}^f \equiv \widehat{c}^f_{\rm V,NC} - c^f_{\rm V,NC}$ with $\widehat{c}^f_{\rm V,NC}$ being the loop-corrected coupling. For definiteness, we take the Fermi constant to be $G^{}_\mu$ as determined precisely from muon decays. The one-loop NC potential is given by $\widehat{\cal V}^\alpha_{\rm NC} = \sqrt{2}G^{}_\mu N^{}_f \widehat{c}^f_{\rm V,NC}$, whereas the tree-level one reads ${\cal V}^{}_{\rm NC} = \sqrt{2}G^{}_\mu N^{}_f c^f_{\rm V,NC}$. In this case, the relative magnitude of one-loop correction to the NC potential is characterized by $\Delta c^f_{\rm V,NC}/c^{f}_{\rm V,NC}$, as $G^{}_\mu$ will be anyway assigned the experimentally measured value in both tree- and loop-level calculations. Similarly, the correction to the CC potential will be represented by $\Delta c^e_{\rm V, CC}/c^e_{\rm V, CC}$, where $\Delta c_{\rm V,CC}^e \equiv \widehat{c}^e_{\rm V,CC} - c^e_{\rm V,CC}$ and $\widehat{c}^e_{\rm V,CC}$ is the loop-level coupling.
	
	\subsection{On-shell Renormalization}
	
	The one-loop renormalization of the SM in the on-shell scheme can be found in the monograph~\cite{Bohm:2001yx} and also in many excellent review papers~\cite{Aoki:1982ed, Bohm:1986rj, Hollik:1988ii, Denner:1991kt}. For completeness, a brief summary of the on-shell renormalization of the SM is presented in Appendix~\ref{app:A}, and the basic procedure is sketched in this subsection in order to explain our conventions.
	
	For the classical Lagrangian of the standard electroweak theory, we shall closely follow the definitions and notations in Ref.~\cite{Denner:1991kt}. As usual, the quantization of the SM can be performed by introducing the gauge-fixing terms and the Faddeev-Popov ghosts, and then the Feynman rules can be derived, where the 't Hooft-Feynman gauge will be chosen for simplicity. At the one-loop level, the ultraviolet (UV) divergences in the one-point Green's function (i.e., the Higgs tadpole diagrams), one-particle-irreducible two-point Green's functions and three-point vertex functions can be separated out by using the dimensional regularization, where the space-time dimension is set to $d = 4 - 2\epsilon$ and the UV-divergent term in the limit of $\epsilon \to 0$ shows up as
	\begin{eqnarray}
		\Delta \equiv \frac{1}{\epsilon} - \gamma_{\rm E}^{} + \ln (4\pi) \; ,
	\end{eqnarray}
	where $\gamma_{\rm E}^{} \approx 0.577$ is the Euler-Mascheroni constant. In principle, only the particle masses and coupling constants need to be renormalized to guarantee finite $S$-matrix elements in the SM~\cite{Sirlin:1977sv, Sirlin:1980nh}, but the wave-function renormalization of physical fields is necessary to keep the Green's functions finite as well.
	
	After expressing the bare model parameters and physical fields in terms of the renormalized ones and the corresponding counterterms, as summarized in Appendix~\ref{app:A}, one can calculate the Higgs tadpole diagrams, two-point Green's functions and three-point vertex functions, which are in general UV-divergent. Then, the on-shell renormalization conditions on the renormalized Green functions are imposed to remove the UV-divergences and thus determine the counterterms. Finally, a complete set of renormalized parameters are chosen as inputs and implemented to calculate the $S$-matrix elements of our interest. Some comments are helpful.
	\begin{itemize}
		\item {\bf Input parameters}. As has been done in Ref.~\cite{Denner:1991kt}, we shall choose the input parameters as the fine structure constant $\alpha$, the $W$-boson mass $m_W^{}$, the $Z$-boson mass $m^{}_Z$, the Higgs-boson mass $m_h^{}$, and the charged-fermion masses $m_f^{}$. Since $m_W^{}$ and $m_Z^{}$ have been chosen as input parameters, the weak mixing angle is defined via $\cos\theta_{\rm w} \equiv m_W^{} / m_Z^{}$. For later convenience, the abbreviations $c \equiv \cos \theta_{\rm w}^{}$ and $s \equiv \sin \theta_{\rm w}^{}$ will be used. Moreover, $s^{}_{2{\rm w}} \equiv \sin2\theta_{\rm w}^{} = 2sc$ and $c_{2{\rm w}}^{} \equiv \cos 2\theta_{\rm w}^{} = c^2 - s^2$ are also implemented to simplify the expressions. 
		
		With the physical parameters chosen above, the electromagnetic coupling constant $e = \sqrt{4\pi \alpha}$ is related to the weak gauge coupling constant $g$ via the weak mixing angle, i.e.,  $e = g s$. Whenever the coupling constants $e$ and $g$ appear, their definitions should be understood in terms of the fine-structure constant $\alpha$ and the weak mixing angle $\theta^{}_{\rm w}$. 
		
		\item {\bf One-loop amplitudes}. The contributions to the amplitudes of $\nu^{}_\alpha + f \to \nu^{}_\alpha + f$ at the one-loop level can be divided into three categories, i.e., the self-energies of weak gauge bosons including the tadpole diagrams, the vertex corrections and the box diagrams. With all the counterterms previously determined in the on-shell scheme, the UV-divergent terms are all canceled out and the finite corrections are obtained. The one-loop diagrams have been calculated by using {\tt Package-X}~\cite{Patel:2015tea,Patel:2016fam}, and the Passarino-Veltman functions~\cite{Passarino:1978jh} are implemented to express one-loop integrals as in Appendix~\ref{app:A}. 
		
		In the following expressions, $x_i^{} \equiv m_i^2/m_W^2$ and $y_i^{} \equiv m_i^2/m_Z^2$ are introduced with ``$i$" referring to the particle type. The fermion masses for external legs are retained, but they are much smaller compared to the gauge-boson masses and thus all the terms of ${\cal O}(x_e^{})$ or ${\cal O}(x_q^{})$ for $q = u,d$ can be safely neglected. It should be noticed that as we are interested in the forward scattering amplitudes, the diagrams with the photon propagator with $p^2 = 0$ attached to the charged fermions will not contribute due to the on-shell renormalization of the electric charge. In addition, neutrinos are massless in the SM and the quark flavor mixing is ignored. For the latter assumption, the reason is simply that the off-diagonal elements of the Cabibbo-Kobayashi-Maskawa (CKM) matrix are much smaller than the diagonal ones and the vertices involving a pair of quarks not in the same isospin-doublet are highly suppressed. 
		
		\item {\bf Finite corrections}. Once the finite corrections to the amplitudes are obtained, one can extract the vector-type coefficients in the corresponding effective Hamiltonian and derive the one-loop corrections to the matter potentials of neutrinos. For the NC part, the renormalized self-energy of the $Z$-boson, the neutrino or charged-fermion vertex, and the box diagrams are denoted as ${\rm i} \Sigma_{Z}^{\rm r}$, ${\rm i} e \Gamma^{\rm r}_{\nu_\alpha \nu_\alpha Z}$ or ${\rm i} e \Gamma^{\rm r}_{ffZ}$ and ${\rm i}{\cal M}_{\rm NC}^f$, respectively, so the correction to the vector-type coupling is 
		\begin{eqnarray}\label{eq:delta_cVNC}
			\Delta c_{\rm V,NC}^f = \left(-\frac{\Sigma_Z^{\rm r}}{m_Z^2} + s_{2{\rm w}}^{} \Gamma_{\nu_\alpha^{} \nu_\alpha^{} Z}^{\rm r} \right) c_{\rm V,NC}^f + s_{2{\rm w}}^{} \Gamma_{ffZ}^{\rm r} - \frac{4m_W^2}{g^2} {\cal M}_{\rm NC}^{f} \;.
		\end{eqnarray}
		Similarly, for the CC part, with the renormalized self-energy of the $W$-boson, the corrected vertex, and the box diagrams denoted as ${\rm i} \Sigma_{W}^{\rm r}$, ${\rm i} e \Gamma^{\rm r}_{\nu_e^{} e W}$ and ${\rm i} {\cal M}_{\rm CC}^{}$, respectively, the correction to the vector-type coupling turns out to be 
		\begin{eqnarray}\label{eq:delta_cVCC}
			\Delta c_{\rm V,CC}^e = \left(-\frac{\Sigma_W^{\rm r}}{m_W^2} + 2\times\sqrt{2} s \Gamma_{\nu_e^{} e W}^{\rm r} \right) c_{\rm V,CC}^e - \frac{4m_W^2}{g^2} {\cal M}_{\rm CC}^{} \;.
		\end{eqnarray} 
		Note that the factor of two associated with the vertex correction $\Gamma_{\nu_e^{} e W}^{\rm r}$ in Eq.~(\ref{eq:delta_cVCC}) arises from the fact that the $\nu^{}_e$-$e$-$W$ vertex appears twice in the diagrams.
	\end{itemize}
	
	The self-energy, vertex and box contributions on the right-hand sides of Eqs.~(\ref{eq:delta_cVNC}) and (\ref{eq:delta_cVCC}) will be presented in Sec.~\ref{sec:VNC} and Sec.~\ref{sec:VCC}, respectively. With the latest values of the input parameters, we shall evaluate these finite corrections in Sec.~\ref{sec:num}.
	
	\section{The Neutral-current Potential}
	\label{sec:VNC}
	\begin{figure}[t]
		\centering
		\includegraphics[scale=1]{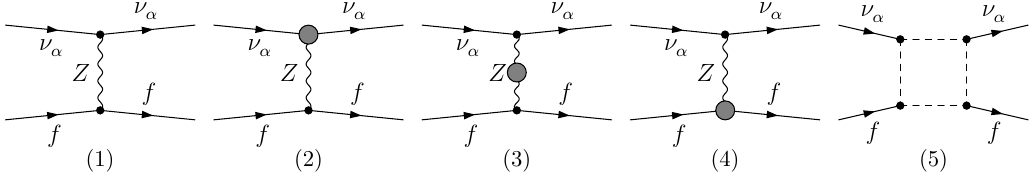} 
		\vspace{-0.2cm}
		\caption{The relevant Feynman diagrams of the scattering $\nu_\alpha^{} + f \to \nu_\alpha^{} + f$ for the NC potential with $f=u,d,e$ at the tree level (1) and at the one-loop level (2)-(5). The shaded circle represents the radiative corrections to the vertices and propagators. The box diagrams in (5) are UV-finite and the dashed box indicates all possible realizations of internal lines.}
		\label{fig:NC}
	\end{figure}
	
	\subsection{The Fermi Constant}
	
	As shown in Eqs.~(\ref{eq:Vl}) and (\ref{eq:Ve}), the NC and CC potentials at the tree level are usually given in terms of the Fermi constant $G^{}_\mu$, which is related to the adopted physical parameters by $G^{}_\mu = g^2/(4\sqrt{2}m^2_W) = \pi \alpha/(\sqrt{2} m^2_W s^2)$. At the one-loop level, however, such a relation is corrected as 
	\begin{eqnarray}\label{eq:Gmuloop}
		\frac{g^2}{4 \sqrt{2} m_W^2} \equiv \widehat{G}_{\mu}^{} \left(1 - \Delta r \right) \;,
	\end{eqnarray}
	where $\widehat{G}_\mu^{}$ stands for the one-loop corrected Fermi constant and the finite radiative corrections are collected in $\Delta r$. With the help of Eqs.~(\ref{eq:tadpole})-(\ref{eq:AZ_mixing}), we can evaluate $\Delta r$ by~\cite{Denner:1991kt} 
	\begin{eqnarray} \label{eq:Delta_r}
		\Delta r  &=& - \left. \frac{\partial \Sigma_{\rm T}^{A}\left(p^{2}\right)}{\partial p^{2}}\right|_{p^{2}=0} + \frac{c^2}{s^2}\left[\frac{\Sigma_{\rm T}^{Z} \left(m_Z^2\right)}{m_Z^2} - \frac{\Sigma_{\rm T}^{W} \left(m_W^2\right)}{m_W^2}\right] + \frac{\Sigma_{\rm T}^W(m_W^2) - \Sigma_{\rm T}^W(0)}{m_W^2} \nonumber \\
		&& - 2\frac{c}{s} \frac{\Sigma_{\rm T}^{AZ} (0)}{m_Z^2} + \frac{\alpha}{4\pi s^2}\left[6+\frac{7-4s^2}{2s^2}\ln\left(\frac{m_W^2}{m_Z^2}\right)\right] \;.
	\end{eqnarray}
	
	Since the Fermi constant determined from the muon lifetime is the most precise, it is convenient to use it in the studies of low-energy weak interactions. For the tree-level matter potential, one may just input the value of $G^{}_\mu$ extracted from the muon lifetime, namely, $G^{}_\mu = G^{\rm exp}_\mu$. On the other hand, at the one-loop level, we implement the relation in Eq.~(\ref{eq:Gmuloop}) to determine $\widehat{G}^{}_\mu$ from the same experimental observation, i.e., $\widehat{G}^{}_\mu (1 - \Delta r) = G^{\rm exp}_\mu$. In this case, the tree-level matter potential is given by ${\cal V} = \sqrt{2}G^{}_\mu N^{}_f c^f_{\rm V}$, while the one-loop potential is $\widehat{\cal V} = \sqrt{2}\widehat{G}^{}_\mu (1 - \Delta r) N^{}_f (c^f_{\rm V} + \Delta c^f_{\rm V})$. As the experimental value $G^{\rm exp}_\mu$ is used to evaluate the matter potential at either the tree- or one-loop level, we shall characterize the magnitude of radiative corrections by
	\begin{eqnarray}\label{eq:1_loop_V}
		\frac{\widehat{\cal V}}{\cal V} - 1 = \frac{\sqrt{2} G_\mu^{\rm exp} N_f^{} \left(c_{\rm V}^f + \Delta c_{\rm V}^f \right)}{\sqrt{2}G^{\rm exp}_\mu N^{}_f c^f_{\rm V}} - 1 = \frac{\Delta c_{\rm V}^f}{c_{\rm V}^f} \;.
	\end{eqnarray}
	It is worthwhile to mention that Eq.~(\ref{eq:1_loop_V}) is applicable to both NC and CC potentials, for which one should make use of the corresponding vector-type couplings and their radiative corrections. Therefore, in the subsequent discussions, we focus only on the radiative corrections to the vector-type couplings.
	
	\subsection{Self-energy of $Z$-boson}
	
	The relevant Feynman diagrams of the scattering $\nu_\alpha^{} + f \to \nu_\alpha^{} + f$ for the NC potential have been shown in Fig.~\ref{fig:NC}. After calculating the one-loop amplitudes, we can extract the corrections to the vector-type coupling $c_{\rm V,NC}^f$. 
	
	First, let us look at the self-energy of $Z$-boson in Fig.~\ref{fig:NC}-(3), where the shaded circle represents all possible contributions. The self-energy of $Z$-boson contributes to $\Delta c_{\rm V,NC}^f$ as $-(c_{\rm V,NC}^f / m_Z^2) \Sigma_Z^{\rm r}$, where ${\rm i} \Sigma_Z^{\rm r}$ denotes the renormalized self-energy.
	
	\begin{itemize}
		\item {\bf Bosonic Contributions}. The bosonic contributions to the $Z$-boson self-energy involve gauge bosons, the Higgs boson, the Goldstone bosons and the Faddeev-Popov ghosts running in the loop. The final result can be written as 
		\begin{eqnarray}
			(4\pi)^2 \Sigma_{Z-{\rm b}}^{\rm r} &=& \frac{g^2 m_Z^2}{8 c^2 \left(1-y_h^{}\right)}\left(y_h^4-6y_h^3+17y_h^2-22y_h^{}+4\right)\ln y_h^{} \nonumber \\  
			&& -\frac{3}{2} g^2 m_Z^2 \left(4 c^4+4 c^2-1\right)  {\rm DiscB}\left(m_Z^2,m_W^{},m^{}_W\right) \nonumber \\
			&& +\frac{g^2 m_Z^2}{4 c^2 \left(y_h^{} - 4 \right)}\left(y_h^3 -7y_h^2 + 20y_h^{} -28\right) {\rm DiscB}\left(m_Z^2,m_h^{},m_Z^{}\right) \nonumber \\
			&& +\frac{g^2 m_Z^2}{24 c^2}\left(6y_h^2 -21y_h^{} -288c^6-264c^4 +112c^2 +49  \right)\;,
		\end{eqnarray}
		where the function ${\rm DiscB}\left(p^2,m_0^{},m_1^{}\right)$ is related to the Passarino-Veltman function via
		\begin{eqnarray}
			{\rm B}^{}_0 \left(p^2;m^{}_0,m^{}_1\right) = \Delta + \ln \left(\frac{\mu ^2}{m_1^2}\right) +2+ {\rm DiscB}(p^2,m^{}_0,m^{}_1)-\frac{m_0^2-m_1^2+p^2}{2 p^2}\ln \left(\frac{m_0^2}{m_1^2}\right)\;, \quad
		\end{eqnarray}
		with $\mu$ being the renormalization mass scale. The explicit form of ${\rm DiscB}\left(p^2,m_0^{},m_1^{}\right)$ reads
		\begin{eqnarray}		
			{\rm DiscB}\left(p^2, m^{}_0, m^{}_1\right) = \frac{\sqrt{\lambda\left(m_0^2,m_1^2,p^2\right)}}{p^2}\ln \left[\frac{m_0^2 + m_1^2 -p^2 + \sqrt{\lambda\left(m_0^2,m_1^2,p^2\right)}}{2m^{}_0 m^{}_1}\right]\;,
		\end{eqnarray}
		where the K\"{a}ll\'{e}n function
		\begin{eqnarray}
			\lambda\left(x,y,z\right) \equiv x^2+y^2+z^2-2xy-2yz-2zx
		\end{eqnarray}
		has been defined. 
		
		\item {\bf Fermionic Contributions}. For the fermions running in the loop, we have
		\begin{eqnarray}
			(4\pi)^2 \Sigma_{Z-{\rm f}}^{\rm r} &=& \sum_f \frac{4 e^2 m_Z^2 }{12 y^{}_f-3}\left\{   6 y^{}_f \left[a_f^2 (1-4 y^{}_f)+2 v_f^2 y^{}_f\right] {\rm DiscB}\left(m_Z^2,m^{}_f,m^{}_f\right) \right. \nonumber \\
			&& \left. +(4 y^{}_f-1) \left[a_f^2 (1-12 y^{}_f)+v_f^2 (6 y^{}_f+1)\right]   \right\}\;,
		\end{eqnarray}
		where we have defined $v_f^{} \equiv c_{\rm V,NC}^f/s^{}_{2{\rm w}}$ and $a_f^{} \equiv c_{\rm A,NC}^f/s^{}_{2{\rm w}}$. Note that the summation is over all the SM fermions and three colors for each type of quarks are taken into account. 
	\end{itemize} 
	
	\subsection{Vertex Contributions}
	
	Then, we calculate the vertex corrections, for which the Feynman diagrams have been depicted in Fig.~\ref{fig:NC}-(2) and (4). For later convenience, we introduce the following functions
	\begin{eqnarray}
		{\cal F}_Z^{} (p^2) &=& \sum_f \left\{  \left[4 a_f^2 m_f^2-p^2\left(a_f^2+v_f^2\right)\right]{\rm B}_0^{}\left(p^2;m_f^{},m_f^{}\right) \right. \nonumber \\
		&&\left. \quad -4 \left(a_f^2+v_f^2\right){\rm B}^{}_{00}\left(p^2;m_f^{},m_f^{}\right) +2 \left(a_f^2+v_f^2\right) {\rm A}^{}_0(m_f^{})   \right\} \;, \\
		{\cal F}_W^{} (p^2) &=& \sum_{\left\{f,f'\right\}} \left[  \left(m_f^2+m_{f^\prime}^2\right) {\rm B}^{}_0\left(p^2;m^{}_f,m^{}_{f'}\right)-4{\rm B}^{}_{00}\left(p^2;m^{}_f,m^{}_{f'}\right) \right. \nonumber \\
		&& \left. \quad -p^2{\rm B}^{}_0\left(p^2;m^{}_f,m^{}_{f'}\right) +{\rm A}^{}_0(m^{}_f)+{\rm A}^{}_0(m^{}_{f'})   \right]\;, \\
		{\cal F}_A^{} (p^2) &=& \sum_f  Q_f^2 \left[  -4 {\rm B}^{}_{00}\left(p^2;m^{}_f,m^{}_f\right)-p^2 {\rm B}^{}_0\left(p^2;m^{}_f,m^{}_f\right) + 2 {\rm A}^{}_0(m^{}_f)  \right] \;, \\
		{\cal F}_{AZ}^{} (p^2) &=& \sum_f Q_f^{} v_f^{} \left[  -4
		{\rm B}^{}_{00}\left(p^2;m^{}_f,m^{}_f\right)-p^2{\rm B}^{}_0\left(m_Z^2;m^{}_f,m^{}_f\right)+2 {\rm A}^{}_0(m^{}_f)  \right] \;,
	\end{eqnarray}
	where $Q^{}_f$ denotes the electric charge and $\{f, f^\prime\}$ refers to the pair of fermions in the same isospin-doublet. As the subscripts of these functions indicate, they represent the contributions from the self-energies of $Z$-boson, $W$-boson, photon and the $A$-$Z$ mixing in Eqs.~(\ref{eq:Z_self-energy})-(\ref{eq:AZ_mixing}). In addition, their derivatives ${\cal F}_{V}^\prime (m_V^2) \equiv \left. {\rm d}{\cal F}_{V}^{}(p^2)/{\rm d}p^2\right|^{}_{p^2=m_V^2}$ for $V = W,Z,A$ are also needed. 
	\begin{itemize}
		\item {\bf The $\nu_\alpha^{}$-$\nu_\alpha^{}$-$Z$ Vertex}. The contribution to $\Delta c_{\rm V,NC}^f$ is given by $s^{}_{2{\rm w}} c_{\rm V,NC}^f \Gamma_{\nu_\alpha^{} \nu_\alpha^{} Z}^{\rm r}$ with
		\begin{eqnarray}
			(4\pi)^2 \Gamma_{\nu_\alpha^{}\nu_\alpha^{} Z}^{\rm r} &=& -\frac{g^2 x_\alpha^{}}{s^{}_{2{\rm w}}}\left(\ln x_\alpha^{} +3\right)  + \frac{ g^2 c^{}_{2{\rm w}}}{s^{}_{2{\rm w}}} \left[ \frac{{\cal F}_Z^{}(m_Z^2)}{m_Z^2} -\frac{{\cal F}_W^{}(m_W^2)}{4 s^2 m_W^2 } \right] + \frac{g^2 s}{2c} \left[{\cal F}_Z^\prime(m_Z^2) - {\cal F}_A^\prime(0)\right] \nonumber\\
			&& +\frac{g^2}{48 c s^{3}}\left(120 c^6+68 c^4-106 c^2+17\right){\rm DiscB}\left(m_Z^2,m^{}_W,m^{}_W\right) \nonumber \\
			&& -\frac{g^2}{6 s^3_{2{\rm w}} \left(y^{}_h-4\right)} \left[  \left(4 c^2-3\right) y_h^3-\left(29 c^2-21\right) y_h^2 \right. \nonumber \\
			&& \left.+\left(88 c^2-60\right) y^{}_h -132 c^2+84  \right] {\rm DiscB}\left(m_Z^2,m^{}_h,m^{}_Z\right)  \nonumber \\
			&& -\frac{g^2 }{48 c^5 s^{3}} \left(96 c^8+88 c^6-100 c^4+14c^2+1\right){\rm DiscB}\left(m_W^2,m^{}_W,m^{}_Z\right)\nonumber \\
			&& + \frac{g^2 c^{}_{2{\rm w}} }{48  c s^3}\left(x_h^2-4 x^{}_h +12 \right) {\rm DiscB}\left(m_W^2,m^{}_h,m^{}_W\right) \nonumber \\
			&& +\frac{g^2 }{12 s_{2{\rm w}}^3 }\left[\left(4c^2-3\right) y_h^3-\left(21 c^2-15\right) y_h^2+\left(42 c^2-30\right) y^{}_h-60 c^2+36\right] \ln y^{}_h \nonumber \\
			&& -\frac{g^2 c^{}_{2{\rm w}} }{12 s^3_{2{\rm w}}}\left(c^2 x_h^3-6 c^2 x_h^2+12 c^2 x^{}_h-24\right)\ln x^{}_h \nonumber \\
			&& +\frac{g^2 }{96 c^7 s^{3} }\left[\left(12 c^6-6 c^4\right) y^{}_h-158 c^6+106 c^4-12 c^2-1 \right]\ln\left(\frac{m_W^2}{m_Z^2}\right)  \nonumber \\
			&& +\frac{g^2 }{48 c^5 s } \left[\left(4 c^2-1\right) y_h^2-6 c^2 y^{}_h-240 c^8-356 c^6+252 c^4+10 c^2-1\right] \;.
		\end{eqnarray}
		Notice that the flavor-dependent terms proportional to $x_\alpha^{}$ are the same as those in Ref.~\cite{Mirizzi:2009td,Botella:1986wy}, and our results are also consistent with Eqs.~(5.46) and (5.47) in Ref.~\cite{Aoki:1982ed}.	
		
		\item {\bf The $f$-$f$-$Z$ Vertex}. With the radiative corrections to the vector-type couplings in the renormalized vertices $\Gamma_{ffZ}^{\rm r}$, the total contributions to $\Delta c_{\rm V,NC}^f $ can be expressed as $s_{2{\rm w}}^{} \Gamma_{ffZ}^{\rm r} $ for $f=u,d,e$. All the terms proportional to the quark and electron masses of ${\cal O}(x_f^{})$ can always be neglected due to the suppression by the $W$-boson mass. 
		\begin{itemize}
			\item \underline{$u$-$u$-$Z$ Vertex}. The renormalized vertex reads 
			\begin{eqnarray}
				(4\pi)^2\Gamma_{uuZ}^{\rm r} &=&  \frac{g^2\left(5-2c^2\right)}{6 s^{}_{2{\rm w}}}\left[\frac{{\cal F}_Z^{}(m_Z^2)}{m_Z^2}  - \frac{{\cal F}_W^{}(m_W^2)}{4 s^{2} m_W^2} \right]  + \frac{e^2 \left(8c^2-5\right)}{6 s^{}_{2{\rm w}}}\left[{\cal F}_Z^\prime(m_Z^2) - {\cal F}_A^\prime(0)\right]  \nonumber \\
				&& +\frac{4 e^2 }{3 m_Z^2} {\cal F}_{AZ}^{} (m_Z^2) -\frac{g^2 \left(2 c^2-5\right)}{288 c s^{3}} \left(x_h^2-4 x_h^{}+12\right) {\rm DiscB}\left(m_W^2,m_h^{},m_W^{}\right) \nonumber \\ 
				&& -\frac{g^2}{36 s^{3}_{2{\rm w}} \left(y_h^{}-4\right)}\left[ \left(16 c^4-28 c^2+15\right) y_h^3 - \left(104 c^4 - 185 c^2+105\right) y_h^2 \right. \nonumber \\
				&& \left. +\left(256 c^4-472 c^2+300\right) y_h^{} -288 c^4+564 c^2-420 \right] {\rm DiscB}\left(m_Z^2,m_h^{},m_Z^{}\right) \nonumber \\
				&& +\frac{g^2 }{96 c s^{3}}\left(320 c^8-360 c^6-236 c^4+398 c^2-23\right){\rm DiscB}\left(m_Z^2,m_W^{},m_W^{}\right) \nonumber \\
				&& + \frac{ g^2}{288 c^5 s^{3}}\left(96 c^8-104 c^6-372 c^4+78 c^2+5\right) {\rm DiscB}\left(m_W^2,m_Z^{},m_W^{}\right) \nonumber \\
				&& +\frac{g^2 }{72 s^{3}_{2{\rm w}}} \left[ \left(16 c^4-28 c^2+15\right) y_h^3 -\left(72 c^4 - 129 c^2 + 75\right) y_h^2 \right. \nonumber \\
				&& \left. +\left(144 c^4-258 c^2+150\right) y_h^{} - 96 c^4+204 c^2-180 \right] \ln y_h^{} \nonumber \\
				&&  + \frac{g^2\left(2 c^2-5\right)  }{72 s^{3}_{2{\rm w}}} \left(c^2 x_h^3-6 c^2 x_h^2+12 c^2 x_h^{}-24\right) \ln x_h^{} \nonumber \\
				&& + \frac{g^2 }{576 c^7 s^{3}}\left[\left(30 c^4-12 c^6\right) y_h^{} + 16 c^8+134 c^6-418 c^4+68 c^2+5\right]\ln\left(\frac{m_W^2}{m_Z^2}\right) \nonumber \\
				&& + \frac{g^2}{288 c^5 s }\left[ \left(16 c^4-12 c^2+5\right) y_h^2 -6 c^2 \left(8 c^2-5\right) y_h^{} \right. \nonumber \\
				&& \left. -1920 c^{10}+400 c^8+1652 c^6-1100 c^4-42 c^2+5  \right] \;.
			\end{eqnarray}
			
			\item \underline{$d$-$d$-$Z$ Vertex}. The renormalized vertex is given by
			\begin{eqnarray}
				(4\pi)^2\Gamma_{ddZ}^{\rm r} &=& - \frac{g^2\left(2c^2+1\right)}{6 s^{}_{2{\rm w}}}\left[\frac{{\cal F}_Z^{}(m_Z^2)}{m_Z^2} - \frac{{\cal F}_W^{}(m_W^2)}{4 s^2 m_W^2 } \right]  + \frac{e^2(1-4c^2)}{6 s^{}_{2{\rm w}}}\left[{\cal F}_Z^\prime(m_Z^2) - {\cal F}_A^\prime(0)\right]  \nonumber \\
				&& -\frac{2 e^2}{3 m_Z^2} {\cal F}_{AZ}^{} (m_Z^2) -\frac{g^2\left(2 c^2+1\right)}{288 c s^{3}}\left(x_h^2-4 x^{}_h+12\right) {\rm DiscB}\left(m_W^2,m^{}_h,m^{}_W\right) \nonumber \\
				&& +\frac{g^2}{36 s^{3}_{2{\rm w}} \left(y^{}_h-4\right)}\left[ \left(8 c^4-8 c^2+3\right) y_h^3-\left(52 c^4-49 c^2+21\right) y_h^2 \right. \nonumber \\
				&& \left.  +\left(128 c^4-104 c^2+60\right) y^{}_h - 144 c^4+84 c^2-84  \right]{\rm DiscB}\left(m_Z^2,m^{}_h,m^{}_Z\right) \nonumber \\
				&& -\frac{g^2 }{96 c s^{3}}\left(160 c^8-120 c^6-84 c^4+146 c^2-3\right){\rm DiscB}\left(m_Z^2,m^{}_W,m^{}_W\right) \nonumber \\
				&& +\frac{g^2 }{288 c^5 s^3}\left(96 c^8+184 c^6+36 c^4-18 c^2-1\right){\rm DiscB}\left(m_W^2,m^{}_Z,m^{}_W\right) \nonumber \\
				&& -\frac{g^2 }{72 s^{3}_{2{\rm w}}} \left[ \left(8 c^4-8 c^2+3\right) y_h^3 - \left(36 c^4 - 33 c^2 + 15\right) y_h^2  \right. \nonumber \\
				&& \left. +\left(72 c^4-66 c^2+30\right) y^{}_h  -48 c^4+12 c^2 -36 \right] \ln y^{}_h \nonumber \\
				&& + \frac{g^2 \left(2 c^2+1\right)}{72 s^3_{2{\rm w}}} \left(c^2 x_h^3-6 c^2 x_h^2+12 c^2 x^{}_h-24\right)\ln x^{}_h \nonumber \\
				&& -\frac{g^2 }{576 c^7 s^3} \left[6 \left(2 c^2+1\right) c^4 y_h + 8 c^8-170 c^6-50 c^4+16 c^2+1\right] \ln\left(\frac{m_W^2}{m_Z^2}\right)\nonumber \\
				&&  -\frac{g^2 }{288 c^5 s }\left[ \left(8 c^4+1\right) y_h^2 + 6 c^2 \left(1-4 c^2\right) y^{}_h \right. \nonumber \\
				&& \left. + \left(960 c^{10}+160 c^8-292 c^6+172 c^4+6 c^2-1\right)  \right]  \;.
			\end{eqnarray}
			
			\item \underline{$e$-$e$-$Z$ Vertex}. The renormalized vertex is 
			\begin{eqnarray}
				(4\pi)^2\Gamma_{eeZ}^{\rm r} &=& \frac{g^2 \left(2 c^2-3\right) }{2 s^{}_{2{\rm w}} }\left[\frac{{\cal F}_Z^{}(m_Z^2)}{m_Z^2} -\frac{{\cal F}_W^{}(m_W^2)}{4 s^{2} m_W^2} \right] + \frac{e^2 \left(3-4 c^2\right)}{2 s^{}_{2{\rm w}}}\left[{\cal F}_Z^\prime(m_Z^2) - {\cal F}_A^\prime(0)\right] \nonumber \\
				&& - \frac{2 e^2}{m_Z^2} {\cal F}_{AZ}^{} (m_Z^2)  +\frac{g^2 \left(2 c^2-3\right)  }{96 c s^{3}} \left(x_h^2-4 x_h^{}+12\right) {\rm DiscB}\left(m_W^2,m_h^{},m_W^{}\right) \nonumber \\
				&& +\frac{g^2}{12 s^3_{2{\rm w}} \left(y_h-4\right)}\left[ \left(8 c^4-16 c^2+9\right) y_h^3-\left(52 c^4-107 c^2+63\right) y_h^2 \right. \nonumber \\
				&& \left. +\left(128 c^4-280 c^2+180\right) y_h^{}-144 c^4+348 c^2-252 \right]{\rm DiscB}\left(m_Z^2,m_h^{},m_Z^{}\right) \nonumber \\
				&& -\frac{g^2 }{96 c s^{3}}\left(480 c^8-600 c^6-388 c^4+650 c^2-43\right){\rm DiscB}\left(m_Z^2,m_W^{},m_W^{}\right) \nonumber \\
				&& -\frac{g^2}{96 c^5 s^{3}}\left(96 c^8-8 c^6-236 c^4+46 c^2+3\right){\rm DiscB}\left(m_W^2,m_W^{},m_Z^{}\right) \nonumber \\
				&& -\frac{g^2 }{24 s^3_{2{\rm w}}} \left[ \left(8 c^4-16 c^2+9\right) y_h^3 - \left(36 c^4 - 75 c^2 + 45\right) y_h^2  \right. \nonumber \\
				&& \left. +\left(72 c^4-150 c^2+90\right) y_h^{}-48 c^4+132 c^2 -108 \right] \ln y_h^{} \nonumber \\
				&& -\frac{g^2 \left(2 c^2-3\right)}{24 s^3_{2{\rm w}}} \left(c^2 x_h^3-6 c^2 x_h^2+12 c^2 x_h^{}-24\right) \ln x_h^{} \nonumber \\
				&& -\frac{g^2}{192 c^7 s^3} \left[ \left(18 c^4-12 c^6\right) y_h^{} + 96 c^{12}-240 c^{10}+224 c^8 \right. \nonumber \\
				&& \left. +62 c^6-250 c^4+40 c^2 +3   \right] \ln\left(\frac{m_W^2}{m_Z^2}\right) \nonumber \\
				&& - \frac{g^2 }{96 c^5 s}\left[ \left(8 c^4-8 c^2+3\right) y_h^2 - 6 c^2 \left(4 c^2-3\right) y_h^{} \right. \nonumber \\
				&& \left. - 960 c^{10}+320 c^8+1004 c^6-676 c^4-26 c^2+3  \right] \;. 
			\end{eqnarray} 
			This renormalized vertex has also been calculated in Ref.~\cite{Aoki:1982ed}, where the results in Eqs.~(5.42)-(5.44) agree perfectly with ours.	
			
		\end{itemize}	
		
	\end{itemize}
	
	\subsection{Box-diagram Contributions}
	
	Finally, we consider the box diagrams shown in Fig~\ref{fig:NC}-(5). The contribution to $\Delta c_{\rm V,NC}^f$ is actually given by $-(4 m_W^2/g^2) {\cal M}^{f}_{\rm NC} $, where the relevant amplitudes from the one-loop box diagrams are expressed as ${\rm i} {\cal M}^{f}_{\rm NC}$ with $f=u,d,e$. These amplitudes are UV-finite and no renormalization is needed. For the scattering with the neutrino $\nu_\alpha^{}$, the box diagrams for three different types of background particles lead to
	\begin{eqnarray}
		(4\pi)^2{\cal M}_{\rm NC}^u &=& -\frac{g^4}{8 m_W^2}\left[\frac{5-4c^2}{4c^2}+ x_\alpha^{} \left(\ln x_\alpha^{} +1\right)\right]\;,  \\
		(4\pi)^2{\cal M}_{\rm NC}^d &=& +\frac{g^4}{2m_W^2}\left[\frac{20 c^2-1}{16 c^2 } +x_\alpha^{}\left(\ln x_\alpha^{}+1\right)\right]\;, \\
		(4\pi)^2{\cal M}_{\rm NC}^e &=& +\frac{g^4}{2m_W^2}\left[\frac{28c^2-9}{16c^2}+x_\alpha^{} \left(\ln x_\alpha^{}+1\right)\right] \;.
	\end{eqnarray}
	The first two results are consistent with Eqs.~(7.1)-(7.3) in Ref.~\cite{Sakakibara:1980hw}, whereas the final one is the same as in Eq.~(5.51) of Ref.~\cite{Aoki:1982ed}. The neutrino flavor-dependent parts have been found to be compatible with the previous calculations in Refs.~\cite{Mirizzi:2009td,Botella:1986wy}. 
	
	\section{The Charged-current Potential}
	\label{sec:VCC}
	
	\begin{figure}[t]
		\centering
		\includegraphics[scale=1]{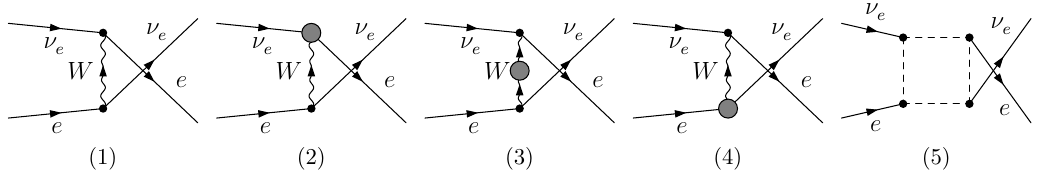}
		\vspace{-0.2cm}
		\caption{The relevant Feynman diagrams of the scattering $\nu_e^{} + e \to \nu_e^{} + e$ for the CC potential at the tree level (1) and at the one-loop level (2)-(5). The notations are the same as those in Fig.~\ref{fig:NC}.} 
		\label{fig:CC}
	\end{figure}
	In parallel with the discussions about the NC potential, there are also three types of radiative corrections to the CC potential ${\cal V}_{\rm CC}^{}$, which will be denoted by $\Delta c_{\rm V,CC}^e$. The relevant Feynman diagrams of the elastic scattering between electron neutrinos and electrons $\nu_e^{} +e \to \nu_e^{} + e$ for the CC potential have been given in Fig.~\ref{fig:CC}. 
	
	\subsection{Self-energy of $W$-boson}
	
	First, we consider the self-energy of $W$-boson in Fig.~\ref{fig:CC}-(3), where the shaded circle represents all possible contributions. The contribution to $\Delta c_{\rm V,CC}^e$ from the $W$-boson self-energy can be expressed as $-(c_{\rm V,CC}^e / m_W^2) \Sigma_W^{\rm r}$, where ${\rm i} \Sigma_W^{\rm r}$ denotes the renormalized self-energy.
	\begin{itemize}
		\item {\bf Bosonic Contributions}. The $W$-boson self-energy receives the contributions from all the bosons running in the loop, and the renormalized self-energy is 
		\begin{eqnarray}
			(4\pi)^2\Sigma_{W-{\rm b}}^{\rm r} &=& -\frac{g^2 m_Z^2}{4 c^2} \left(12 c^6+44 c^4-13 c^2-1\right){\rm DiscB}\left(m_W^2,m_Z^{},m_W^{}\right) \nonumber \\
			&& +\frac{g^2 m_W^2}{4\left(x_h-4\right)}\left(x_h^3-7 x_h^2+20 x_h^{} -28 \right){\rm DiscB}\left(m_W^2,m_h^{},m_W^{}\right) \nonumber \\
			&& -\frac{g^2 m_W^2 }{8 (x_h^{}-1)} \left(x_h^4-6 x_h^3+17 x_h^2-22 x_h^{}+4\right) \ln x_h^{} \nonumber \\
			&& - \frac{g^2 m_Z^2}{8 c^4 s^2}\left(16 c^{10}-4 c^8-118 c^6+83 c^4-10 c^2-1\right) \ln\left(\frac{m_W^2}{m_Z^2}\right)  \nonumber \\
			&& + \frac{g^2 m_W^2 }{24 c^4} \left[c^4 \left(6 x_h^2-21 x_h^{} -370\right)+75 c^2+6\right] \;.
		\end{eqnarray}
		
		\item {\bf Fermionic Contributions}. The self-energy correction with fermions in the loop reads
		\begin{eqnarray}
			(4\pi)^2\Sigma_{W-{\rm f}}^{\rm r} &=& g^2 m_W^2\sum_{\{f,f^\prime\}} \left\{\frac{m_W^4}{6} \left\{\frac{-x_f^3+x_f^2 x_{f^\prime}^{}+x_f^{} \left(x_{f^\prime}^2-4 x_{f^\prime}^{} +3 \right)-x_{f^\prime}^3+3 x_{f^\prime}^{} -2 }{\lambda \left(m_f^2,m_{f^\prime}^2,m_W^2\right)} \right. \right. \nonumber \\
			&& \left. -\frac{1}{m_W^4}\left[3 x_f^2-2x_f^{} \left(3 x_{f^\prime}^{}-1\right)+3 x_{f^\prime}^2+2 x_{f^\prime}^{}-2\right] \right\} {\rm DiscB}\left(m_W^2,m_f^{},m_{f^\prime}^{}\right) \nonumber \\
			&& + \frac{1}{4\left(x_f^{}-x_{f^\prime}^{} \right)}\left[x_f^4-4 x_f^3 x_{f^\prime}^{} +x_f^2 \left(6 x_{f^\prime}^2-1\right)-4 x_f^{} x_{f^\prime}^3+x_{f^\prime}^4-x_{f^\prime}^2\right] \ln \left(\frac{x_f^{}}{x_{f^\prime}^{}}\right) \nonumber \\
			&& \left. -\frac{1}{12}\left[6 x_f^2+3 x_f^{} \left(1-4 x_{f^\prime}^{}\right)+6 x_{f^{\prime}}^2+3 x_{f^{\prime}}^{}-4 \right] \right\}\;.
		\end{eqnarray}
		We should sum over all the contributions from the SM fermions, where $\{f, f^\prime\}$ denotes the pair of fermions in the same isospin-doublet, and take into account three colors for each type of quarks. 
	\end{itemize}
	
	\subsection{Vertex Contributions}
	Then, we turn to the CC vertex corrections, which have been shown in Fig.~\ref{fig:CC}-(2) and (4). The total contribution to $\Delta c_{\rm V,CC}^e$ from the $\nu_e^{}$-$e$-$W$ vertex can be expressed as $\sqrt{2} s \Gamma_{\nu_e^{} e W}^{\rm r} c_{\rm V,CC}^e$ with the renormalized vertex ${\rm i} \Gamma_{\nu_e^{} e W}^{\rm r}$ defined as follows
	\begin{eqnarray}
		(4\pi)^2\Gamma_{\nu_e^{} e W}^{\rm r} &=& \frac{g^2}{c^2}\left[\frac{{\cal F}_Z^{}(m_Z^2)}{m_Z^2} - \frac{{\cal F}_W^{}(m_W^2)}{4 s^2 m_W^2}\right] - e^2 \left[{\cal F}_A^\prime(0) - \frac{{\cal F}_W^\prime(m_W^2)}{4 s^2} \right] \nonumber \\
		&& +\frac{g^2 }{24  s^2\left(4-x_h^{}\right)} \left[ \left(c^2-2\right) x_h^3 - \left(5 c^2-13\right) x_h^2 \right. \nonumber \\
		&& \left. +4 \left(c^2-8\right) x_h^{} +12 \left(c^2+3\right)  \right]{\rm DiscB}\left(m_W^2,m_h^{},m_W^{} \right)  \nonumber \\
		&& -\frac{g^2}{24 c^4 s^2 }\left(60 c^8-8 c^6+71 c^4-22 c^2-2\right)  {\rm DiscB}\left(m_W^2,m_W^{},m_Z^{}\right) \nonumber \\
		&& - \frac{g^2}{24  s^2}\left(y_h^2-4 y_h^{}+12\right) {\rm DiscB}\left(m_Z^2,m_h^{},m_Z^{}\right) \nonumber \\
		&& +\frac{g^2}{24 s^2 }\left(48 c^6+68 c^4-16 c^2-1\right)  {\rm DiscB}\left(m_Z^2,m_W^{},m_W^{}\right) \nonumber \\
		&& +\frac{g^2}{48 s^2} \left(y_h^3-6 y_h^2 +18 y_h^{} -20 c^2 \right)\ln y_h^{} \nonumber \\
		&& -\frac{g^2}{48} \left[\left(c^4+c^2+2\right) x_h^3 - \left(6 c^2+9\right) x_h^2 +18 x_h^{} +168 c^2-8\right] \ln x_h^{} \nonumber \\
		&& +\frac{g^2}{48 c^6 s^2} \left( c^6 y_h^3 -6 c^6 y_h^2 +18 c^6 y_h^{} - 48 c^{10}-36 c^8 \right. \nonumber \\
		&& \left. +166 c^6-119 c^4+18 c^2+2 \right)\ln \left(\frac{m_W^2}{m_Z^2}\right) \nonumber \\
		&& +\frac{g^2}{24 c^4}\left[\left(c^2+2\right) y_h^2-6 c^2 y_h^{} -96 c^8-224 c^6+32 c^4+23 c^2+2 \right] \;.
	\end{eqnarray}
	As mentioned before, the same CC vertex appears both in Fig.~\ref{fig:CC}-(2) and (4), so a factor of two is present in the vertex correction in Eq.~(\ref{eq:delta_cVCC}).
	
	\subsection{Box-diagram Contributions}
	
	Finally, the contributions from the UV-finite box diagrams should be included, for which the Feynman diagram has been shown in Fig.~\ref{fig:CC}-(5). Since the electrons are present in the background, electron neutrinos interact with them via both NC and CC processes. In particular, for the box diagrams, it is impossible to categorize the contributions into either NC or CC type. However, it is clear that both $\nu^{}_\mu$ and $\nu^{}_\tau$ interact with the background particles only through the NC interaction. For this reason, we select the box diagrams that are universal for all three types of neutrinos as the NC part, whereas the remaining ones as the CC part. The contribution from box diagrams can be written as $-(4 m_W^2/g^2) {\cal M}^{}_{\rm CC}$ with the amplitude
	\begin{eqnarray}
		(4\pi)^2 {\cal M}^{}_{\rm CC} = -\frac{g^4}{8 m_W^2 s^2}\left[2 s^4 \left(\ln x_e^{}-1\right)+\left(2 c^4+6 c^2-3\right) \ln \left(\frac{m_W^2}{m_Z^2}\right)\right] \;.
	\end{eqnarray}
	Here it is worth mentioning that for the box diagram involving the internal photon propagator, the generalized Fierz identity~\cite{Nieves:2003in}
	\begin{eqnarray}
		\overline{\nu_e^{} (x)} \left(1+\gamma^5\right) e(x)\ \overline{e(x)} \left(1-\gamma^5\right) \nu_e^{} (x) = -\frac{1}{2} \overline{\nu_e^{} (x)} \gamma_\mu \left(1-\gamma^5\right) \nu_e^{} (x)\ \overline{e(x)} \gamma^\mu \left(1+\gamma^5\right) e(x) \; ,
	\end{eqnarray}
	has been utilized to transform the contributions into the correction to the vector-type coupling.
	
	\section{Numerical Results}
	\label{sec:num}
	
	Given the finite corrections in the previous sections, we now specify the input parameters and evaluate the one-loop corrections to the matter potentials. The latest values of relevant input parameters are quoted from the Particle Data Group~\cite{ParticleDataGroup:2022pth} and summarized below:
	\begin{itemize}
		\item The fine structure constant
		\begin{eqnarray}
			\alpha \equiv e^2/(4\pi) = 1/137.035999084 \; ;
		\end{eqnarray}
		
		\item The gauge-boson and Higgs-boson masses\footnote{The latest measurement of $W$-boson mass given by the CDF-II collaboration is $m_W^{} = 80.433~{\rm GeV}$~\cite{CDF:2022hxs}, yielding a $7\sigma$ discrepancy with the SM expectation. However, we have checked that the difference in the correction to the matter potential caused by such a discrepancy appears at the order of ${\cal O}(10^{-4})$.}
		\begin{eqnarray}
			m_W^{} = 80.377~{\rm GeV}\;, \qquad \ m_Z^{} = 91.1876~{\rm GeV} \;, \quad \ m_h^{} = 125.25~{\rm GeV}\;;
		\end{eqnarray}
		
		\item The quark masses
		\begin{eqnarray}
			&& m_u^{} = 2.16~{\rm MeV} \;, \qquad m_c^{} = 1.67~{\rm GeV} \;, \qquad m_t^{} = 172.5~{\rm GeV} \;, \nonumber\\
			&& m_d^{} = 4.67~{\rm MeV} \;, \qquad m_s^{} = 93.4~{\rm MeV} \;, \qquad m^{}_b = 4.78~{\rm GeV} \;;
		\end{eqnarray}
		
		\item The charged-lepton masses
		\begin{eqnarray}
			m_e^{} = 0.511~{\rm MeV} \;, \qquad m_\mu^{} = 105.658~{\rm MeV} \;, \qquad m_\tau^{}= 1.777~{\rm GeV} \;.
		\end{eqnarray}
	\end{itemize}
	
	All the particle masses quoted above refer to the on-shell masses, except for those of three light quarks (i.e., $u$, $d$ and $s$). Instead, the running masses of three light quarks at the energy scale of $\mu = 2~{\rm GeV}$ are used, since the on-shell masses of light quarks are not well-defined due to the non-perturbative nature of quantum chromodynamics at low energies.
	
	\begin{table}[t]
		\centering \begin{spacing}{1.15}
			\begin{tabular}{cccccc}
				\hline\hline
				& Self-energy & $\nu_\alpha^{}$-$\nu_\alpha^{}$-$Z$ &  $f$-$f$-$Z$ & Box diagrams & $\Delta c_{\rm V,NC}^f$ \\ 
				\hline
				\multirow{2}{*}{$f=u$} & \multirow{2}{*}{$-2.1 \times 10^{-3}$} & $5.1 \times 10^{-3}$ & \multirow{2}{*}{$-6.0 \times 10^{-3}$} & $7.9 \times 10^{-4}$ & \multirow{2}{*}{$-2.2 \times 10^{-3}$} \\
				& & $1.5 \times 10^{-6}$ (fd) & & $-4.2 \times 10^{-6}$ (fd) &   \\
				\hline
				\multirow{2}{*}{$f=d$} & \multirow{2}{*}{$3.7 \times 10^{-3}$} & $-8.8 \times 10^{-3}$ & \multirow{2}{*}{$-3.3 \times 10^{-3}$} & $-6.1 \times 10^{-3}$ & \multirow{2}{*}{$-1.5 \times 10^{-2}$} \\
				& & $-2.6 \times 10^{-6}$ (fd) & & $1.7 \times 10^{-5}$ (fd) &   \\
				\hline
				\multirow{2}{*}{$f=e$} & \multirow{2}{*}{$5.6 \times 10^{-4}$} & $-1.4 \times 10^{-3}$ & \multirow{2}{*}{$15.3 \times 10^{-3}$} & $-5.3 \times 10^{-3}$ & \multirow{2}{*}{$9.2 \times 10^{-3}$} \\
				& & $-3.9 \times 10^{-7}$ (fd) & & $1.7 \times 10^{-5}$ (fd) &   \\
				\hline\hline	
		\end{tabular}\end{spacing}
  \vspace{0.2cm}
		\caption{The one-loop corrections to the vector-type couplings $c_{\rm V,NC}^f$ for $f=u,d,e$ from the $Z$-boson self-energy, vertex corrections and box diagrams. Here ``fd" stands for the flavor-dependent part, for which we choose the heaviest charged lepton $\alpha=\tau$ as an example.}	
		\label{tab:NC}
	\end{table}
	
	From Eq.~(\ref{eq:Vl}), we can observe that the tree-level NC potential induced by each type of fermions in the matter is proportional to the vector-type coupling $c_{\rm V,NC}^u = 0.2026$, $c_{\rm V,NC}^d = -0.3514$ and $c_{\rm V,NC}^e = -0.0539$, where these couplings have been displayed in Table~\ref{tab:cv_ca} and evaluated by using $s^2 = 1 - m^2_W/m^2_Z \approx 0.223$. The corresponding corrections to these vector-type couplings from the $Z$-boson self-energy, vertex corrections and box diagrams are listed in Table~\ref{tab:NC}, accordingly. The flavor-dependent corrections are labeled as ``fd", where we have chosen the flavor $\alpha = \tau$ for example. It shows clearly that the flavor-dependent contributions are two to three orders of magnitude smaller than the flavor-independent ones. Therefore, in the final results of $\Delta c_{\rm V,NC}^f$ in the last column of Table~\ref{tab:NC}, we only list the dominant flavor-independent values.    
	
	Then, we can translate the NC potential induced by quarks and electrons into that by protons, neutrons and electrons via the relations among their number densities, namely, $N_u^{} = 2 N_p^{} + N_n^{}$, $N_d^{} = N_p^{} + 2 N_n^{}$ and $N_e^{} = N_p^{}$. The one-loop correction to the NC potential is thus given by
	\begin{eqnarray}
		\frac{\Delta c_{\rm V,NC}^{}}{c_{\rm V,NC}^{}} = \frac{N_p^{} \left(2\Delta c_{\rm V,NC}^u + \Delta c_{\rm V,NC}^d + \Delta c_{\rm V,NC}^e\right) + N_n^{} \left(\Delta c_{\rm V,NC}^u + 2 \Delta c_{\rm V,NC}^d\right)}{N_n^{} \left( c_{\rm V,NC}^u + 2 c_{\rm V,NC}^d\right)} \approx 0.062 + 0.02 \frac{N^{}_p}{N^{}_n}\;, \qquad
	\end{eqnarray}
	where the relation $2 c_{\rm V,NC}^u +  c_{\rm V,NC}^d + c_{\rm V,NC}^e = 0$ has been implemented. Therefore, for the ordinary matter with $N^{}_p \approx N^{}_n$, the one-loop correction to the NC potential is about $8.2\%$.
	
	\begin{table}[t]
		\centering \begin{spacing}{1.3}
			\begin{tabular}{cccc}
				\hline\hline 
				Self-energy & $\nu_e^{}$-$ e $-$W$ & box diagrams &  $\Delta c_{\rm V,CC}^e$  \\ 
				\hline
				$-6.4 \times 10^{-3}$ & $4.5 \times 10^{-2}$ & $1.9 \times 10^{-2}$ & $5.8 \times 10^{-2}$ \\ 
				\hline\hline	
		\end{tabular}\end{spacing}
  \vspace{0.2cm}
		\caption{The one-loop contributions to $\Delta c_{\rm V,CC}^e $ from the $W$-boson self-energy, vertices and box diagrams. The tree-level vector-type coupling is $c^e_{\rm V, CC} = 1$.}	
		\label{tab:CC}
	\end{table}
	Similar to the case of the NC potential, we collect all the contributions to $\Delta c_{\rm V,CC}^e$ in Table~\ref{tab:CC}. It shows that there is a correction of about $6\%$ to the CC matter potential. Whereas the NC potentials are the same for three-flavor neutrinos, except for the tiny flavor-dependent contributions, this correction to the CC potential of electron neutrinos will play an important role in neutrino flavor conversions. In the near future, the long-baseline accelerator neutrino experiments DUNE and T2HK will make use of the MSW effect to resolve the sign of $\Delta m^2_{31}$, and also determine the octant of $\theta_{23}^{}$ and the CP-violating phase $\delta_{\rm CP}^{}$. In consideration of both one-loop corrections to the matter potential and the uncertainty in the matter density, we shall carry out a more dedicated study to explore their impact on the determination of neutrino mass ordering and the precise measurements of the CP-violating phase at both DUNE and T2HK in a separate work. 

\section{Results in the $\overline{\rm MS}$ scheme}
\label{sec:msbar}

Although we have carried out all the calculations in the on-shell scheme, it is also interesting to make a comparison with those in the $\overline{\rm MS}$ scheme. For the one-loop corrections to the MSW potentials in the $\overline{\rm MS}$ scheme, the relevant Feynman diagrams are the same as in the on-shell scheme. But some differences should be noted. First, the input parameters are chosen as the running fine-structure constant $\alpha(\mu)$ and the running masses $\{m_W^{}(\mu), m_Z^{}(\mu),m_h^{}(\mu)$, $m_f^{}(\mu)\}$. The bare parameters and fields, marked by the subscript ``0", can be separated into the renormalized ones and the renormalization constants. For the coupling and mass parameters, we have
    \begin{eqnarray} \label{eq:MSbar_mass_function}
    	e_0^{} = Z_e^{\overline{\rm MS}} e(\mu) = \left(1+\delta Z_e^{\overline{\rm MS}}\right)e(\mu)\;, \quad M^2_0 = M_{}^2 (\mu) + \delta M_{\overline{\rm MS}}^2 \;, 
    \end{eqnarray}
where $M$ represents $m_W^{}, m_Z^{},m_h^{}$ and $m_f^{}$, and $\delta M_{\overline{\rm MS}}^2$ is the corresponding counterterm in the $\overline{\rm MS}$ scheme. The bare fields of physical particles can be expressed as follows
	\begin{eqnarray} \label{eq:MSbar_wave_function}
		W_{0\mu}^{\pm} &=& \sqrt{Z_W^{\overline{\rm MS}}} W_\mu^\pm = \left(1 + \frac{1}{2} \delta Z_W^{\overline{\rm MS}}\right) W_\mu^\pm \;, \nonumber \\
		\begin{pmatrix} Z_{0\mu}^{} \\ A_{0\mu}^{}\end{pmatrix} &=& \begin{pmatrix}
			\sqrt{Z_{ZZ}^{\overline{\rm MS}}} & \sqrt{Z_{ZA}^{\overline{\rm MS}}} \\ \sqrt{Z_{AZ}^{\overline{\rm MS}}} & \sqrt{Z_{AA}^{\overline{\rm MS}}} 
		\end{pmatrix}
		\begin{pmatrix} Z_{\mu}^{} \\ A_{\mu}{} \end{pmatrix} = \begin{pmatrix}
			1+ \displaystyle \frac{1}{2} \delta Z_{ZZ}^{\overline{\rm MS}} & \displaystyle \frac{1}{2} \delta Z_{ZA}^{\overline{\rm MS}} \\ \displaystyle  \frac{1}{2} \delta Z_{AZ}^{\overline{\rm MS}} & 1 + \displaystyle \frac{1}{2} \delta Z_{AA}^{\overline{\rm MS}}
		\end{pmatrix}
		\begin{pmatrix} Z_{\mu}^{} \\ A_{\mu}^{} \end{pmatrix}\;, \nonumber \\
		h_0^{} &=& \sqrt{Z_h^{\overline{\rm MS}}} h = \left(1 + \frac{1}{2} \delta Z_h^{\overline{\rm MS}}\right) h\;, \nonumber \\
		f_{i,0}^{} &=& \sqrt{Z_{ij}^{f,\overline{\rm MS}}} f_j^{} = \left(1+\frac{1}{2} \delta Z_{ij}^{f,\overline{\rm MS}}\right) f_{j}^{} \;.
	\end{eqnarray}
Second, different from the on-shell scheme where all the counterterms are fixed by the on-shell renormalization conditions, those in the $\overline{\rm MS}$ scheme contain only the UV-divergent parts. More specifically, after separating out the divergences in the Green's functions, the counterterms are just given by the terms proportional to $\Delta$. For example, with the self-energy correction of $Z$-boson in Fig.~\ref{fig:z}, the wave-function and mass counterterms of $Z$-boson in Eqs.~(\ref{eq:MSbar_mass_function}) and (\ref{eq:MSbar_wave_function}) read
	\begin{eqnarray}
		\delta Z_Z^{\overline{\rm MS}} &=& \left[\frac{g^2 \left(18 c^4+2 c^2-1\right) }{6 c^2  } - \frac{4 e^2 }{3 } \sum_f \left(a^2_f+v^2_f\right)\right]\Delta \;, \nonumber \\ 
		\delta m_{Z,\overline{\rm MS}}^2 &=& -\left\{ \frac{g^2 m_Z^2 \left(42 c^4-10 c^2-7\right) }{6 c^2 } + \frac{4 e^2 }{3 } \sum_f  \left[a_f^2 \left(6 m_f^2- m_Z^2\right)-m_Z^2 v^2_f\right]\right\} \Delta\;.
	\end{eqnarray}
In the above equation, the dependence on the 't Hooft mass scale $\mu$ of the gauge-coupling and masses is suppressed, and the running weak mixing angle is defined as $\cos\theta_{\rm w}^{\overline{\rm MS}} (\mu) \equiv m_W^{}(\mu)/m_Z^{}(\mu)$. The summation over all the SM fermions is implied. After the renormalization procedure in the $\overline{\rm MS}$ scheme, we are left with finite corrections to the scattering amplitudes, from which the vector-type couplings can be extracted and the one-loop corrections to the MSW potential of neutrinos can be derived.
    
We take the neutrino-electron scattering $\nu_\alpha^{} + e \to \nu_\alpha^{} + e $ as an illustrative example, since the background electrons contribute to both NC and CC potentials. The scattering of neutrinos with background quarks can be dealt with in a similar way. Now the scattering amplitudes under consideration are still described by one-loop diagrams in Fig.~\ref{fig:NC} and Fig.~\ref{fig:CC}. In the $\overline{\rm MS}$ scheme, the finite amplitudes expressed in terms of running parameters can be obtained by simply removing the divergent terms. Then, the matter potentials can be extracted from the amplitudes with the help of Eqs.~(\ref{eq:delta_cVNC}) and (\ref{eq:delta_cVCC}). Besides, as the residue of the fermion self-energy is not fixed at 1 in the $\overline{\rm MS}$ scheme, loop corrections to the neutrino and electron external legs should also be taken into account. Such contributions can be treated as connecting the self-energies in the diagrams (1)-(6) of Fig.~\ref{fig:fermion} to the tree-level diagram of the scattering process in Fig.~\ref{fig:NC} and Fig.~\ref{fig:CC}. As shown in Eq.~(\ref{eq:MSbar_wave_function}), the wave-function renormalization of external fermion legs will contribute a factor of $1/2$ to the scattering amplitudes. Collecting all the contributions together, we find the total correction to the vector-type coupling $c_{\rm V,NC}^{e,\overline{\rm MS}}$ as follows
    \begin{eqnarray} \label{eq:delta_cVNCe_MS}
		\Delta c_{\rm V,NC}^{e,\overline{\rm MS}} (\mu) &=& \frac{ 3 g^2 r_t^{} }{4 c^2} \left(4 c^2-3\right) \ln \left(\frac{\mu^2}{m_t^2}\right) -\frac{ 3 g^2 r_h^{} \left(4 c^2-3\right)  }{8 c^2 (r_h^{}-1)} \ln \left(\frac{\mu^2}{m_h^2}\right) \nonumber \\
		&& + g^2 \left(4 c^4-8 c^2+3\right) \ln \left(\frac{\mu^2}{m_W^2}\right) -\frac{g^2 (r_h^{}-4)}{8 c^2 (r_h^{}-1)}\left(4 c^2-3\right) \ln \left(\frac{\mu^2}{m_Z^2}\right) \nonumber \\
		&& + \frac{g^2}{16 c^2}\left[4 c^2 (r_h^{}-21)-3 r_h^{}+39\right] \;.
    \end{eqnarray}
Here we set all the fermions to be massless except for the top quark. For the CC potential, the total correction to the vector-type coupling $c_{\rm V,CC}^{e,\overline{\rm MS}}$ reads
    \begin{eqnarray} \label{eq:delta_cVCCe_MS}
	   \Delta c_{\rm V,CC}^{e,\overline{\rm MS}} (\mu) &=& \frac{3 g^2 r_h^{} }{4 \left(r_h^{}-c^2\right)} \ln \left(\frac{\mu^2}{m_h^2}\right) - \frac{3 g^2 r_t^{}}{2 c^2}\ln \left(\frac{\mu^2}{m_t^2}\right) - \frac{g^2}{s_{\rm 2w}^2 }  \left(7 c^2-4\right) \ln \left(\frac{\mu^2}{m_Z^2}\right) \nonumber \\
        && +\frac{g^2 }{4 s^2 \left(r_h^{} -c^2\right)} \left[8 c^4-c^2 (5 r_h^{}+11)+8 r_h^{}\right] \ln \left(\frac{\mu^2}{m_W^2}\right)  \nonumber \\
        && +\frac{g^2 }{8 c^2} \left(14 c^2-r_h^{}-6 r_t^{}+7\right) \;.
    \end{eqnarray}
The corrections to the NC potential from quarks in the $\overline{\rm MS}$ scheme can also be calculated similarly, and we just give the final results
	\begin{eqnarray} \label{eq:delta_cVNCu_MS}
		\Delta c_{\rm V,NC}^{u,\overline{\rm MS}} (\mu) &=& -\frac{g^2 r_h^{} \left(8 s^2-3\right)}{8 c^2 (r_h^{}-1)}\ln \left(\frac{\mu^2}{m_h^2}\right) +\frac{g^2 r_t^{} \left(8 s^2-3\right)}{4 c^2} \ln \left(\frac{\mu^2}{m_t^2}\right) \nonumber \\
		&& +\frac{1}{3} g^2 \left(3 c^2-8 s^4+3 s^2\right) \ln \left(\frac{\mu^2}{m_W^2}\right)-\frac{g^2 (r_h^{}-4) \left(8 s^2-3\right)}{24 c^2 (r_h^{}-1)} \ln \left(\frac{\mu^2}{m_Z^2}\right) \nonumber \\
		&& +\frac{g^2 }{48 c^2} \left[24 c^2+(r_h^{}-1) \left(8 s^2-3\right)\right] \;,
	\end{eqnarray}
for the up quark;
	\begin{eqnarray} \label{eq:delta_cVNCd_MS}
		\Delta c_{\rm V,NC}^{d,\overline{\rm MS}} (\mu) &=& \frac{g^2 r_h^{} \left(4 s^2-3\right) }{8 c^2 (r_h^{}-1)}\ln \left(\frac{\mu^2}{m_h^2}\right) -\frac{g^2 r_t^{} \left(4 s^2-3\right)}{4 c^2}\ln \left(\frac{\mu^2}{m_t^2}\right) \nonumber \\
		&& -\frac{1}{3} g^2 \left(3 c^2-4 s^4+3 s^2\right) \ln \left(\frac{\mu ^2}{m_W^2}\right) + \frac{g^2 (r_h^{}-4) \left(4 s^2-3\right)}{24 c^2 (r_h^{}-1)} \ln \left(\frac{\mu^2}{m_Z^2}\right) \nonumber \\
		&& +\frac{g^2}{48 c^2} \left[4 c^2 (r_h^{}-37)-r_h^{}+13\right] \;.
	\end{eqnarray}
for the down quark.

The one-loop corrections to the effective couplings have been previously calculated in the low-energy effective theory of the SM in Ref.~\cite{Hill:2019xqk}. The vector-type couplings in our work are related to the lepton coefficients in Ref.~\cite{Hill:2019xqk} by $\widehat{c}_{\rm V,NC}^{e,\overline{\rm MS}} (\mu) = g^{}_{\rm L}(\mu) + g^{}_{\rm R}(\mu)$ and $\widehat{c}_{\rm V,CC}^{e,\overline{\rm MS}} (\mu) = g(\mu)$. For the couplings of quarks, the corresponding relation reads $\widehat{c}_{\rm V,NC}^{q,\overline{\rm MS}} = g^{q}_{\rm L}(\mu) + g^{q}_{\rm R}(\mu)$. Since we are interested in the neutrino forward scattering, there is no contribution from the neutrino-photon coupling $g^{\nu_\ell^{} \gamma}_{}$ in Eq.~(1) of Ref.~\cite{Hill:2019xqk}. On the other hand, as the quark flavor mixing has been ignored, the term associated with $\left|V_{tq}^{}\right|^2$ in $g_{\rm L}^q(\mu)$ is also absent in the final result. With these two points in mind, we have made a comparison between our results and those in Eqs.~(10) and (11) of Ref.~\cite{Hill:2019xqk} and found complete agreement.

As the vector-type couplings have been obtained in both the on-shell and $\overline{\rm MS}$ schemes, it is helpful to compare between the one-loop matter potentials $\widehat{{\cal V}}_{\rm CC}^{}$ and $\widehat{{\cal V}}_{\rm NC}^{}$ in the on-shell scheme and their counterparts $\widehat{{\cal V}}_{\rm CC}^{\overline{\rm MS}}$ and $\widehat{{\cal V}}_{\rm NC}^{\overline{\rm MS}}$ at $\mu = m^{}_Z$ in the $\overline{\rm MS}$ scheme. The matter potentials are physical observables and thus should be scheme-independent. As already shown in Sec.~\ref{sec:num}, with the input on-shell parameters, the CC potential $\widehat{\cal V}_{\rm CC}^{}$ reads
\begin{eqnarray}
    \widehat{\cal V}_{\rm CC}^{} = \frac{\pi\alpha}{m_W^2 s^2} N_e^{} \widehat{c}_{\rm V,CC}^e \approx 1.679\times 10^{-5}\left(\frac{N_e^{}}{{\rm cm}^{-3}}\right)~{\rm GeV}^{-2}~{\rm cm}^{-3}\;,
    \label{eq:Vcc-os}
\end{eqnarray}
where $1~{\rm GeV}^{-2}~{\rm cm}^{-3} \approx 7.684 \times 10^{-33}~{\rm eV}$ in natural units should be noticed. The NC potential from electrons is
\begin{eqnarray}
    \widehat{\cal V}_{\rm NC}^{} = \frac{\pi\alpha}{m_W^2 s^2} N_e^{} \widehat{c}_{\rm V,NC}^e \approx -7.118 \times 10^{-7}\left(\frac{N_e^{}}{{\rm cm}^{-3}}\right)~{\rm GeV}^{-2}~{\rm cm}^{-3}\;.
    \label{eq:Vnc-os}
\end{eqnarray} 
If the input on-shell parameters in our calculations are implemented to extract the $\overline{\rm MS}$ parameters at $\mu = m^{}_Z$ through the matching conditions at one loop order, then one should obtain the same results for the matter potentials. However, to further compare with the results in Ref.~\cite{Hill:2019xqk}, we adopt the input $\overline{\rm MS}$ parameters therein, i.e.,
\begin{eqnarray}
& \alpha^{(5)}(m_Z^{})^{-1} = 127.955\;, \quad  s^2(m_Z^{})\equiv 1- m_W^2(m_Z^{}) / m_Z^2(m_Z^{}) = 0.23144\;, \nonumber \\
& m_W^{}(m_Z^{}) = 80.961~{\rm GeV}\;, \quad m_Z^{}(m_Z^{}) = 92.3499~{\rm GeV}\;,\quad m_t^{}(m_Z^{}) = 170.9~{\rm GeV} \;.
\end{eqnarray}
Since the Higgs boson is absent in the tree-level diagrams, the conversion between the Higgs-boson pole mass to its $\overline{\rm MS}$ mass leads only to higher-order effects and thus its pole mass is used. The effective couplings at the scale $\mu=m_Z^{}$ are listed in the first line of Table~3 in Ref.~\cite{Hill:2019xqk}, which can be converted into $\widehat{c}^{e,\overline{\rm MS}}_{\rm V, CC}(\mu)$ and $\widehat{c}^{e,\overline{\rm MS}}_{\rm V, NC}(\mu)$ at $\mu = m^{}_Z$. Then the matter potentials from electrons in the $\overline{\rm MS}$ scheme can be calculated as below
\begin{eqnarray}
    \widehat{\cal V}_{\rm CC}^{\overline{\rm MS}} = \frac{\pi\alpha^{(5)}_{}(m_Z^{})}{m_W^2(m_Z^{}) s^2(m_Z^{})} N_e^{} \widehat{c}_{\rm V,CC}^{e,\overline{\rm MS}} (m_Z^{}) \approx 1.652\times 10^{-5}\left(\frac{N_e^{}}{{\rm cm}^{-3}}\right)~{\rm GeV}^{-2}~{\rm cm}^{-3}
    \label{eq:Vcc-ms}
\end{eqnarray}
for the CC potential, and
\begin{eqnarray}
    \widehat{\cal V}_{\rm NC}^{\overline{\rm MS}} = \frac{\pi\alpha^{(5)}_{}(m_Z^{})}{m_W^2(m_Z^{}) s^2(m_Z^{})} N_e^{} \widehat{c}_{\rm V,NC}^{e,\overline{\rm MS}} (m_Z^{}) \approx -7.109 \times 10^{-7}\left(\frac{N_e^{}}{{\rm cm}^{-3}}\right)~{\rm GeV}^{-2}~{\rm cm}^{-3}
    \label{eq:Vnc-ms}
\end{eqnarray}
for the NC potential. Compared with the corresponding values in Eqs.~(\ref{eq:Vcc-os}) and (\ref{eq:Vnc-os}) in the on-shell scheme, we find that they are very close to each other. The relative difference for the CC potential is about $1.61\%$ while that for the NC potential is about $0.12\%$. Such a small discrepancy can be attributed to the higher-order corrections that have also been considered in Ref.~\cite{Hill:2019xqk} in the extraction of the input $\overline{\rm MS}$ parameters from the experimental measurements of the on-shell parameters. First, the determination of ${\alpha^{(5)}(m^{}_Z)}^{-1}$ from the fine-structure constant $\alpha$ in the Thomson limit involves the decoupling of quarks and charged leptons and the renormalization-group running. The ${\cal O}(\alpha \alpha_s^{})$ corrections in the beta function of $\alpha(\mu)$ have also been included. Then, the higher-order corrections to the relationship between on-shell and $\overline{\rm MS}$ parameters, such as weak gauge boson masses and the top-quark mass, have been partially considered.

It is worth commenting very briefly on the renormalization schemes for precision calculations. Generally speaking, the on-shell scheme is advantageous in the sense that the on-shell parameters can be directly extracted from experimental measurements. The experimental determination of different $\overline{\rm MS}$ parameters is usually carried out at different energy scales associated with relevant physical processes, so the renormalization-group equations should be implemented to obtain the complete set of $\overline{\rm MS}$ parameters at a common scale. However, the $\overline{\rm MS}$ scheme together with the approach of effective field theories is practically more convenient to deal with higher-order corrections beyond one-loop and any theories with different mass scales, as in Ref.~\cite{Hill:2019xqk}. In any case, our calculations of the one-loop matter potentials for neutrinos in the on-shell scheme are complementary to those in the $\overline{\rm MS}$ scheme in Ref.~\cite{Hill:2019xqk}.

\section{Summary}
\label{sec:sum}
	
In this paper, we have performed a complete calculation of the MSW matter potential for all-flavor neutrinos at the one-loop level in the SM. Following the on-shell renormalization of the SM, we have calculated the one-loop amplitudes for the coherent forward scattering of neutrinos with the SM fermions present in the ordinary matter. The radiative corrections to the vector-type couplings of neutrinos in both NC and CC processes have been obtained and used to determine the MSW matter potential. The same calculations are also carried out in the $\overline{\rm MS}$ scheme, and the results are compared with those in the literature. With the latest values of the SM parameters, we evaluate the finite corrections to the matter potentials and find that the correction to the NC potential is about $8\%$ while that to the CC potential is about $6\%$. 
	
In the coming precision era of neutrino oscillation physics, one has to reconsider the radiative corrections at the percent level to the interactions of neutrinos with matter. For instance, the JUNO experiment will push the relative errors in the measurement of the oscillation parameters $\sin^2\theta^{}_{12}$, $\Delta m^2_{21}$ and $\Delta m^2_{31}$ even down to the sub-percent level~\cite{JUNO:2015zny,JUNO:2022mxj}. The next-generation long-baseline accelerator neutrino experiments are expected to determine the neutrino mass ordering, the octant of $\theta_{23}^{}$ and the value of the CP-violating phase $\delta_{\rm CP}^{}$. The experimental sensitivities of DUNE and T2HK to these unknown parameters are also sufficiently high to probe the one-loop corrections to the MSW matter potential. In this sense, we believe that our calculations are not only useful for the study of neutrino oscillation phenomenology, but also serve as an instructive example for precision calculations in the whole field of neutrino physics.
	
\section*{Acknowledgements}
This work was supported by the National Natural Science Foundation of China under grant No. 11835013. One of the authors (J.H.) would like to thank Dr. Di Zhang for helpful suggestions on using {\tt FeynArts}. All Feynman diagrams in this work are generated by {\tt FeynArts}~\cite{Hahn:2000kx}, and the loop integrals are calculated with the help of {\tt Package-X}~\cite{Patel:2015tea,Patel:2016fam}. 
	
	\appendix
	
	\section{Renormalization of the Standard Model}
	\label{app:A}
	
	In this appendix, we explain some details about the on-shell renormalization of the Standard Model (SM) and list all the relevant one-loop diagrams for completeness.
	
	The renormalization procedure that we have adopted follows closely that in Ref.~\cite{Denner:1991kt}. Instead of repeating the derivations of all the counterterms, we just highlight some key points relevant to our calculations. More details of the on-shell renormalization can be found in a number of excellent reviews~\cite{Aoki:1982ed,Bohm:1986rj,Hollik:1988ii,Denner:1991kt}, where the SM Lagrangian and the Feynman rules are explicitly given.
	
	\subsection{Renormalization Constants}
	
	Once the set of input physical parameters is chosen, one can decompose the bare parameters and fields, which will be marked by the subscript ``0", into the renormalized ones and the counterterms. More explicitly, the bare parameters are given by
	\begin{eqnarray} \label{eq:ct_parameter}
		e_0^{} &=& Z_e^{} e = \left(1 + \delta Z_e^{}\right) e\;, \nonumber\\
		m_{W,0}^2 &=& m_W^2 + \delta m_W^2 \;, \nonumber\\
		m_{Z,0}^2 &=& m_Z^2 + \delta m_Z^2 \;, \nonumber\\
		m_{h,0}^2 &=& m_h^2 + \delta m_h^2 \;, \nonumber\\
		m_{f,0}^2 &=& m_f^2 + \delta m_f^2 \;, 
	\end{eqnarray}
	while the renormalization of the physical fields is as follows
	\begin{eqnarray}\label{eq:ct_field}
		W_{0\mu}^{\pm} &=& \sqrt{Z_W^{}} W_\mu^\pm = \left(1 + \frac{1}{2} \delta Z_W^{}\right) W_\mu^\pm \;, \nonumber \\
		\begin{pmatrix} Z_{0\mu}^{} \\ A_{0\mu}^{}\end{pmatrix} &=& \begin{pmatrix}
			\sqrt{Z_{ZZ}^{}} & \sqrt{Z_{ZA}^{}} \\ \sqrt{Z_{AZ}^{}} & \sqrt{Z_{AA}^{}} 
		\end{pmatrix}
		\begin{pmatrix} Z_{\mu}^{} \\ A_{\mu}{} \end{pmatrix} = \begin{pmatrix}
			1+ \displaystyle \frac{1}{2} \delta Z_{ZZ}^{} & \displaystyle \frac{1}{2} \delta Z_{ZA}^{} \\ \displaystyle  \frac{1}{2} \delta Z_{AZ}^{} & 1 + \displaystyle \frac{1}{2} \delta Z_{AA}^{}
		\end{pmatrix}
		\begin{pmatrix} Z_{\mu}^{} \\ A_{\mu}^{} \end{pmatrix}\;, \nonumber \\
		h_0^{} &=& \sqrt{Z_h^{}} h = \left(1 + \frac{1}{2} \delta Z_h^{}\right) h\;, \nonumber \\
		f_{i,0}^{\rm L} &=& \sqrt{Z_{ij}^{f,{\rm L}}} f_j^{\rm L} = \left(1+\frac{1}{2} \delta Z_{ij}^{f,{\rm L}}\right) f_j^{\rm L} \;, \nonumber \\
		f_{i,0}^{\rm R} &=& \sqrt{Z_{ij}^{f,{\rm R}}} f_j^{\rm R} = \left(1+\frac{1}{2} \delta Z_{ij}^{f,{\rm R}}\right) f_j^{\rm R} \;.
	\end{eqnarray}
	The subscripts $i$ and $j$ of the fermion fields refer to different generations. In our calculations, the flavor mixing among different generations of quarks plays an insignificant role, so we ignore it and its radiative corrections. Hence only the $i=j$ case is considered and the CKM matrix is taken to be the identity matrix. A more careful treatment of the renormalization of the CKM matrix can be found in Refs.~\cite{Denner:1990yz,Gambino:1998ec,Pilaftsis:2002nc}. In addition, the renormalization of unphysical fields is irrelevant to the one-loop scattering amplitudes and will be neglected as well.
	
	\subsection{Fixing the Counterterms}
	
	The one-loop self-energies of the scalar and fermion fields are denoted as ${\rm i}\Sigma$, while those of gauge fields as ${\rm i}\Sigma^{}_{\rm T}$ with 
	\begin{eqnarray}
		{\rm i}\Sigma_{\mu\nu}^{V} (p^2) = {\rm i}\Sigma_{\rm T}^{V} \left(g_{\mu\nu}^{} - \frac{p_\mu^{} p_\nu^{}}{p^2}\right) + {\rm i}\Sigma_{\rm L}^{V} \frac{p_\mu^{} p^{}_\nu}{p^2} \;,
	\end{eqnarray}
	for $V = W, Z, A, AZ$. The counterterms are fixed by imposing the on-shell conditions and can be expressed in terms of the self-energies. The mass and wave-function counterterms of gauge bosons and the Higgs boson are given by
	\begin{eqnarray}\label{eq:gauge_higgs_ct}
		\delta m_W^2 &=& -{\rm Re}\;\Sigma_{\rm T}^W \left(m_W^2\right)\;, \qquad \delta Z_W^{} = \left.{\rm Re}\;\frac{\partial \Sigma_{\rm T}^W \left(p^2\right)}{\partial p^2}\right|_{p^2_{} = m_W^2}\;, \nonumber \\
		\delta m_Z^2 &=& -{\rm Re}\;\Sigma_{\rm T}^Z \left(m_Z^2\right)\; , \qquad \delta Z_Z^{} = \left.{\rm Re}\;\frac{\partial \Sigma_{\rm T}^Z \left(p^2\right)}{\partial p^2}\right|_{p^2_{} = m_Z^2}\;, \nonumber \\
		\delta m_h^2 &=& + {\rm Re}\;\Sigma_{}^h \left(m_h^2\right)\;, \qquad \delta Z_h^{} = - \left.{\rm Re}\;\frac{\partial \Sigma_{}^h \left(p^2\right)}{\partial p^2}\right|_{p^2_{} = m_h^2}\;.
	\end{eqnarray}
	The counterterms for the photon and $A$-$Z$ mixing are
	\begin{eqnarray}\label{eq:gamma_Z_ct}
		\delta Z_{AA}^{} = \left.\frac{\partial \Sigma_{\rm T}^{AA}\left(p^2\right)}{\partial p^2}\right|_{p^2 = 0} \;, \qquad 
		\delta Z_{AZ}^{} = 2 {\rm Re}\;\frac{\Sigma_{\rm T}^{AZ}\left(m_{Z}^{2}\right)}{m_{Z}^{2}}\;, \qquad \delta Z_{Z A}^{} = - 2 \frac{\Sigma_{\rm T}^{A Z}(0)}{m_{Z}^{2}}\;.
	\end{eqnarray}
	Notice that there is a minus sign for the gauge-boson self-energy in our notations compared to those in Refs.~\cite{Bohm:1986rj,Hollik:1988ii,Denner:1991kt}. Such a difference just arises from the definition of the gauge-boson self-energy, which is denoted as ${\rm i}\Sigma_{\rm T}^{}$ in our work while as $-{\rm i}\Sigma_{\rm T}^{}$ in the previous literature. As a result, all the counterterms corresponding to the gauge-boson self-energies in Eqs.~(\ref{eq:gauge_higgs_ct}) and (\ref{eq:gamma_Z_ct}) have an opposite sign.
	
	For the fermion masses and wave functions, the counterterms are fixed by
	\begin{eqnarray}\label{eq:fermion_ct}
		\delta m_f^{} &=& \frac{m_f^{}}{2} {\rm Re} \left[\Sigma^{f,{\rm L}}_{ii}\left(m_f^2\right) + \Sigma^{f,{\rm R}}_{ii}\left(m_f^2\right) + 2 \Sigma^{f,{\rm S}}_{ii} \left(m_f^2\right)\right]\;, \nonumber \\
		\delta Z_{ii}^{f,{\rm L}} &=& - {\rm Re}\;\Sigma_{i i}^{f,{\rm L}}\left(m_{f}^{2}\right) - \left.m_{f}^{2} \frac{\partial}{\partial p^{2}} {\rm Re}\left[\Sigma_{ii}^{f,{\rm L}}\left(p^{2}\right) + \Sigma_{ii}^{f,{\rm R}}\left(p^{2}\right) + 2 \Sigma_{ii}^{f,{\rm S}}\left(p^{2}\right)\right]\right|_{p^{2}=m_{f}^{2}} \;, \nonumber \\
		\delta Z_{ii}^{f,{\rm R}} &=& - {\rm Re}\;\Sigma_{i i}^{f,{\rm R}}\left(m_{f}^{2}\right) - \left.m_{f}^{2} \frac{\partial}{\partial p^{2}} {\rm Re}\left[\Sigma_{ii}^{f,{\rm L}}\left(p^{2}\right) + \Sigma_{ii}^{f,{\rm R}}\left(p^{2}\right) + 2 \Sigma_{ii}^{f,{\rm S}}\left(p^{2}\right)\right]\right|_{p^{2}=m_{f}^{2}} \;.
	\end{eqnarray}
	As has been mentioned in the main text, the terms of ${\cal O}\left(x^{}_f\right)$ can be safely neglected, so only the first terms in the wave-function counterterms of fermions need to be taken into account. Note that the fermion self-energy has been decomposed as below 
	\begin{eqnarray}\label{eq:fermion_self_energy_decomposition}
		\Sigma_{ii}^{f} \left(p^2\right) = \slashed{p} P_{\rm L}^{} \Sigma_{ii}^{f,{\rm L}} \left(p^2\right) + \slashed{p} P_{\rm R}^{} \Sigma_{ii}^{f,{\rm R}}\left(p^2\right) + m^{}_f \Sigma_{ii}^{f,{\rm S}}\left(p^2\right) \;,
	\end{eqnarray}
	with the chiral projection operators $P_{{\rm L},{\rm R}}^{} = \left(1\mp \gamma^5_{}\right)/2$.
	
	The renormalization constant of the electric charge can be expressed in terms of the self-energies by implementing the Ward identity, namely, 
	\begin{eqnarray}\label{eq:charge_ct}
		\delta Z_e^{} = -\frac{1}{2} \delta Z_{AA}^{} - \frac{s}{2c} \delta Z_{ZA}^{} \;,
	\end{eqnarray}
	which is independent of the fermion species. This occurs as the consequence of the universality of the electric charge.
	
	Finally, although the weak mixing angle has not been chosen as an input parameter, it is usually convenient to introduce a counterterm for it as well and use it to simplify the Feynman rules of the vertex counterterms. However, the counterterms of the cosine and sine of the weak mixing angle are related to the counterterms of gauge-boson masses by
	\begin{eqnarray}\label{eq:theta_w_ct}
		\frac{\delta c}{c} = \frac{1}{2} \left(\frac{\delta m_W^2}{m_W^2} - \frac{\delta m_Z^2}{m_Z^2}\right) \; , \qquad \frac{\delta s}{s} = - \frac{c^2}{2s^2} \left(\frac{\delta m_W^2}{m_W^2} - \frac{\delta m_Z^2}{m_Z^2}\right) \;.
	\end{eqnarray}
	
	\subsection{Self-energies}
	
	As all the relevant counterterms are governed by the self-energies, we shall explicitly show the results of the self-energies and give some explanations whenever necessary. In our calculations, the tadpole contribution to the gauge-boson self-energies is included. In subsequent discussions, we focus only on the real parts of the transverse self-energies that contribute to the counterterms.
	
	\subsubsection{Tadpole}
	\begin{figure}[t]
		\centering
		\includegraphics[scale=1]{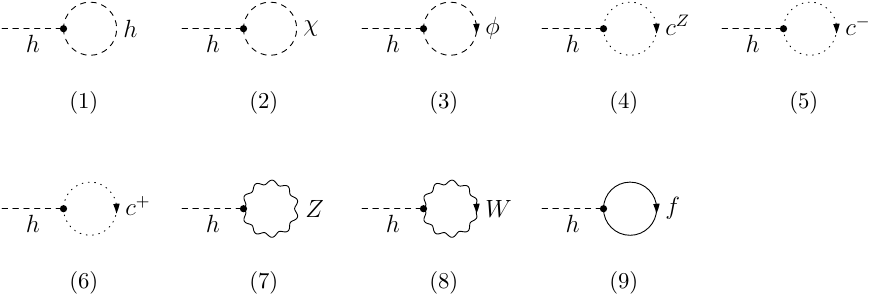}
		\caption{The tadpole diagrams contributing to the gauge-boson self-energies, where the Higgs-boson $h$, the Goldstone bosons $\{\phi^\pm, \chi\}$, the Faddeev-Popov ghosts $\{c^+_{},c^-_{},c^Z_{},c^A_{}\}$, the gauge bosons $\{W^\pm, Z\}$ and all the massive fermions $f$ are running in the loop.}
		\label{fig:tadpole}
	\end{figure}
	The inclusion of the tadpole diagrams ${\rm i}T$ renders the mass counterterms of gauge bosons to be gauge-independent. All the tadpole diagrams are plotted in Fig.~\ref{fig:tadpole}, and the total contribution is
	\begin{eqnarray}\label{eq:tadpole}
		{\rm i} T &=& \frac{{\rm i} g}{(4\pi)^2 4 m_W^{}}\left[ -8 m_f^2 {\rm A}_0^{}(m_f^{})+2 m_h^2
		{\rm A}_0^{}(m_W^{})+m_h^2 {\rm A}_0^{}(m_Z^{})+3
		m_h^2 {\rm A}_0^{}(m_h^{}) \right. \nonumber \\
		&& \left. + 4 d m_W^2
		{\rm A}_0^{}(m_W^{}) -4 m_W^2 {\rm A}_0^{}(m_W^{})+2
		d m_Z^2 {\rm A}_0^{} (m_Z^{})-2 m_Z^2
		{\rm A}_0^{} (m_Z^{}) \right]\;.
	\end{eqnarray}
	Notice that a symmetry factor of $1/2$ should be considered in Fig.~\ref{fig:tadpole}-(1), -(2) and -(7), while a minus sign for the ghost loops in the diagrams (4)-(6) and the fermion loop in the diagram (9) must be included.
	
	\subsubsection{$Z$-boson}
	\begin{figure}[t]
		\centering
		\includegraphics[scale=1]{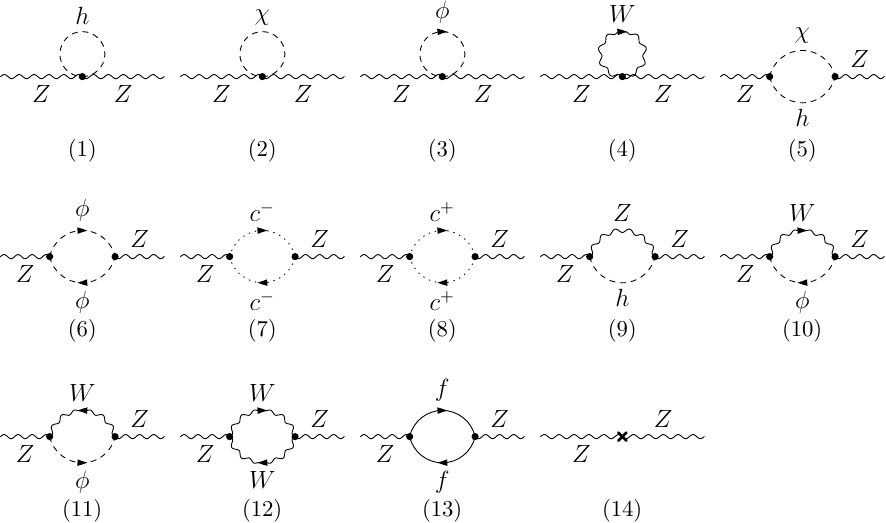}
		\caption{The one-loop diagrams for the $Z$-boson self-energy and the corresponding counterterm in the diagram (14). The notations are the same as those in Fig.~\ref{fig:tadpole}.}
		\label{fig:z}
	\end{figure}
	The one-loop self-energy corrections for $Z$-boson are shown in Fig.~\ref{fig:z}. The contribution to the self-energy of $Z$-boson is 
	\begin{eqnarray}\label{eq:Z_self-energy}
		\Sigma^{Z}_{\rm T} (p^2) &=&  \frac{g^2 }{(4\pi)^2 4 c^2} \left\{  \left(16 c^4 p^2+8c_{2{\rm w}^{}} m_W^2\right) {\rm B}_0^{} \left(p^2;m_W^{},m_W^{}\right)-4m_Z^2 {\rm B}_0^{} \left(p^2;m_Z^{},m_h^{}\right) \right. \nonumber \\ 
		&& +4{\rm B}_{00}^{} \left(p^2;m_h^{},m_Z^{}\right)+\left[16 c^4 (d-1)-16 c^2+4\right] {\rm B}_{00}^{}\left(p^2;m_W^{},m_W^{}\right) -{\rm A}_0^{}(m_h^{}) \nonumber \\
		&& \left. -{\rm A}_0^{}(m_Z^{}) -\left[8 c^4 (d-1)-8 c^2+2\right] {\rm A}^{}_0(m_W^{}) \right\} \nonumber \\
		&& + \frac{2 e^2}{(4\pi)^2} \sum_{f} \left\{ \left[4 a_f^2 m_f^2-p^2 \left(a_f^2+v_f^2\right)\right] {\rm B}_0^{}\left(p^2;m_f^{},m_f^{}\right) \right. \nonumber \\
		&& \left. -4 \left(a^2_f+v^2_f\right) {\rm B}_{00}^{} \left(p^2;m_f^{},m^{}_f\right) +2 \left(a^2_f+v^2_f\right) {\rm A}_0^{}(m^{}_f)  \right\} \;.
	\end{eqnarray}
	The summation is over all the SM fermions. In addition, the tadpole diagrams contribute the term of $g m_Z^{} T/\left(m_h^2 c\right)$.
	
	\subsubsection{$W$-boson}
	\begin{figure}[t]
		\centering
		\includegraphics[scale=1]{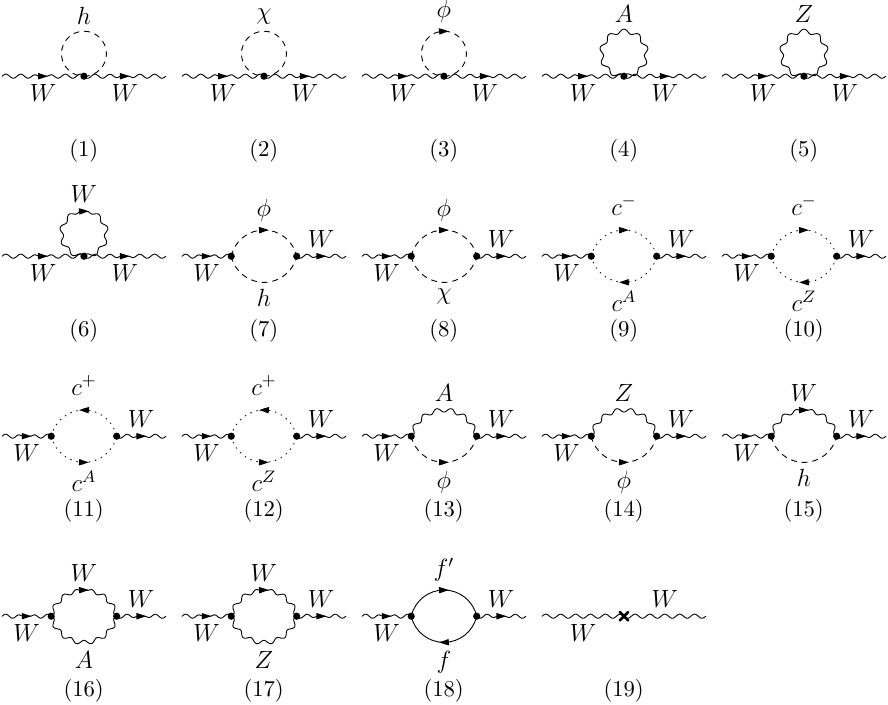}
		\caption{The one-loop diagrams for the $W$-boson self-energy and the corresponding counterterm in the diagram (19).}
		\label{fig:w}
	\end{figure}
	The one-loop diagrams for the $W$-boson self-energy are listed in Fig.~\ref{fig:w}. The total result is 
	\begin{eqnarray}\label{eq:W_self-energy}
		\Sigma_{\rm T}^W \left(p^2\right) &=& \frac{g^2}{4(4\pi)^2} \left\{  \left(8 m_W^2 - 4m_Z^2 s^2 + 16 p^2 c^2\right) {\rm B}_0^{}\left(p^2;m^{}_W,m^{}_Z\right)-4m_W^2{\rm B}_0^{}\left(p^2;m^{}_W,m_h^{}\right)  \right. \nonumber \\
		&&  + 4\left[4 c^2 (d-2) + 1\right] {\rm B}_{00}^{}\left(p^2;m_W^{},m_Z^{}\right)+4{\rm B}_{00}^{}\left(p^2;m_W^{},m_h^{}\right) \nonumber \\
		&& + 4s^2\left(\lambda ^2+4 p^2\right) {\rm B}_0^{}\left(p^2;m_W^{},\lambda \right)+16 s^2 (d-2) {\rm B}_{00}^{}\left(p^2;m_W^{},\lambda \right) \nonumber \\
		&& \left. +(6-4 d) {\rm A}_0^{}(m_W^{}) -4(d-2) s^2 {\rm A}_0^{}(\lambda)-{\rm A}_0^{}(m_h^{}) +\left[4c^2(2-d)-1\right] {\rm A}_0^{}(m_Z^{}) \right\} \nonumber \\
		&& + \frac{g^2}{2(4\pi)^2} \sum_{\left\{f,f'\right\}} \left[ \left(m_f^2 +m_{f^\prime}^{2}\right) {\rm B}_0^{}\left(p^2;m_f^{},m_{f^\prime}^{}\right)-4 {\rm B}_{00}^{}\left(p^2;m_f^{},m^{}_{f^\prime}\right) \right. \nonumber \\
		&& \left. -p^2 {\rm B}_0^{}\left(p^2;m_f^{},m_{f^\prime}^{}\right)+{\rm A}_0^{}(m_f^{})+{\rm A}_0^{}(m_{f^\prime}^{})  \right] \;.
	\end{eqnarray}
	To avoid the infrared divergence, we have introduced a tiny mass $\lambda$ for the photon, which should be kept during the whole calculation and then set to zero in the end. The summation is performed over $\{f,f^\prime\}$, which denotes a pair of fermions in the same isospin-doublet. The tadpole contribution is given by $g m_W^{} T/ m_h^2$.

	\subsubsection{Photon and $A$-$Z$ Mixing}
	\begin{figure}[t]
		\centering
		\includegraphics[scale=1]{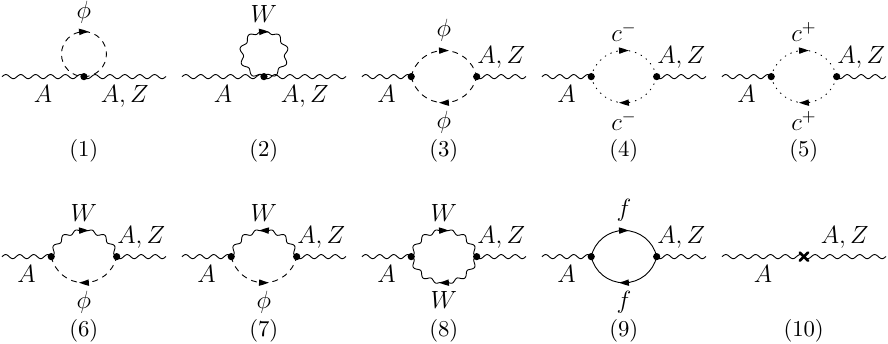}
		\caption{The one-loop diagrams for the self-energy of the photon and the $A$-$Z$ mixing, and the corresponding counterterms in the diagram (10).}
		\label{fig:A}
	\end{figure}
	Different from the cases of gauge bosons, whose self-energies directly contribute to the corrections of the matter potential, the self-energy of the photon and the $A$-$Z$ mixing are relevant for the counterterm of the electric charge as indicated in Eq.~(\ref{eq:charge_ct}). Given one-loop diagrams in Fig.~\ref{fig:A}, the self-energy of the photon $A$ reads 
	\begin{eqnarray}\label{eq:A_self-energy}
		\Sigma_{\rm T}^A \left(p^2\right) &=& \frac{2 e^2}{(4\pi)^2} \left[ \left(3p^2 + 4m_W^2\right) {\rm B}_0^{}\left(p^2;m_W^{},m_W^{}\right)-2(d-2)  {\rm A}_0^{}(m_W^{})\right] \nonumber \\
		&& + \frac{2 e^2}{(4\pi)^2} \sum_f Q_f^2 \left[-4 {\rm B}_{00}^{}\left(p^2;m_f^{},m_f^{}\right)-p^2 {\rm B}_0^{}\left(p^2;m_f^{},m_f^{}\right)+2 {\rm A}_0^{}(m_f^{})\right] \;.
	\end{eqnarray}
	Note that there is no correction to the longitudinal self-energy of the photon, as expected from the unbroken U(1) gauge symmetry.
	
	The Feynman diagrams for the $A$-$Z$ mixing are similar to those for the photon self-energy, as shown in Fig.~\ref{fig:A}. The analytical expression reads 
	\begin{eqnarray}\label{eq:AZ_mixing}
		\Sigma_{\rm T}^{AZ} \left(p^2\right) &=& \frac{g^2 s }{(4\pi)^2 c}\left\{ \left[c^2 (3-2d)+s^2\right]{\rm A}_0^{}(m_W^{}) +2\left[c^2 (2 d-3)-s^2\right] {\rm B}_{00}^{}\left(p^2;m_W^{},m_W^{}\right) \right.  \nonumber \\
		&&   \left. + 2\left[c^2 \left(m_W^2+2 p^2\right)+m_W^2 s^2\right] {\rm B}_0^{}\left(p^2;m_W^{},m_W^{}\right) \right\}  \nonumber \\
		&& + \frac{2 e^2}{(4\pi)^2} \sum_f Q_f^{} v_f^{} \left[-4 {\rm B}_{00}^{}\left(p^2;m_f^{},m_f^{}\right)-p^2 {\rm B}_0^{}\left(p^2;m_f^{},m_f^{}\right) +2 {\rm A}_0^{}(m_f^{})\right] \;.
	\end{eqnarray}
	As in the case of the photon self-energy, the diagrams with the ghost loops and those with the $W$-$\phi$ loops give identical corrections.
	
	\subsubsection{Fermion}
	
	\begin{figure}[t]
		\centering
		\includegraphics{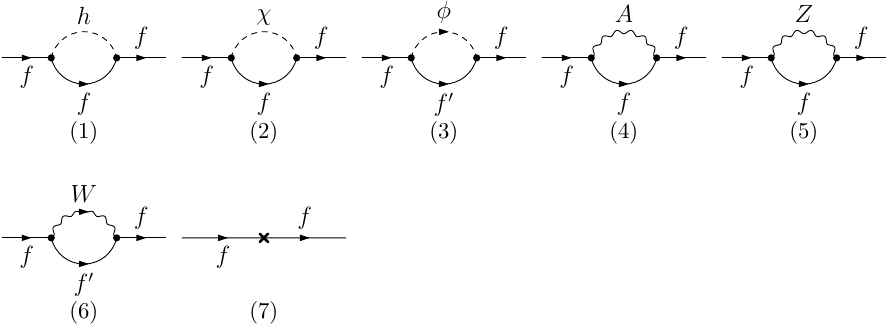}
		\caption{The one-loop diagrams for the self-energy of the fermion $f$ and the corresponding counterterm in the diagram (7). Here $f^\prime$ and $f$ represent the fermions in the same isospin-doublet.}
		\label{fig:fermion}
	\end{figure}
	
	The fermion self-energy will be involved in the vertex counterterms. From the one-loop diagrams in Fig.~\ref{fig:fermion} and with the decomposition in Eq.~(\ref{eq:fermion_self_energy_decomposition}), we obtain 
	\begin{eqnarray}\label{eq:fermion_self-energy}
		\Sigma^{f,{\rm L}} (p^2) &=& -\frac{g^2}{4(4\pi)^2} \left\{ \left[4 (d-2) s^2 (a_f^{}+v_f^{})^2+x_f^{}\right] {\rm B}_1\left(p^2;m_f^{},m_Z^{}\right)+x_f^{} {\rm B}_1\left(p^2;m_f^{},m_h^{}\right) \right. \nonumber \\
		&& \left.  +4 (d-2) Q_f^2 s^2 {\rm B}_1\left(p^2;m_f^{},\lambda \right)+2 (d+x_{f^\prime}^{}-2) {\rm B}_1\left(p^2;m_{f^\prime}^{},m_W^{}\right) \right\} \;, \nonumber \\
		\Sigma^{f,{\rm R}} (p^2) &=& -\frac{g^2}{4(4\pi)^2} \left\{ 4 (d-2) s^2 \left[(a_f^{}-v_f^{})^2 {\rm B}_1\left(p^2;m_f^{},m_Z^{}\right)+Q_f^2 {\rm B}_1\left(p^2;m_f^{},\lambda \right)\right] \right. \nonumber \\
		&& + \left. x_f^{} \left[{\rm B}_1\left(p^2;m_f^{},m_h^{}\right)+{\rm B}_1\left(p^2;m_f^{},m_Z^{}\right)+2 {\rm B}_1\left(p^2;m_{f^\prime}^{},m_W^{}\right)\right] \right\} \;, \nonumber \\
		\Sigma^{f,{\rm S}} (p^2) &=& \frac{g^2}{4(4\pi)^2} \left\{ \left[4 s^2 d \left(a_f^2-v_f^2\right)-x_f^{}
		\right]{\rm B}_0^{}\left(p^2;m_f^{},m_Z^{}\right) \right. \nonumber \\
		&& \left. -2 \left[2 d Q_f^2 s^2 {\rm B}_0^{}\left(p^2;m_f^{},\lambda\right)+x_{f^\prime}^{} {\rm B}_0^{}\left(p^2;m_{f^\prime}^{},m_W^{}\right)\right]+x_f^{} {\rm B}_0^{}\left(p^2;m_f^{},m_h^{}\right) \right\} \;.
	\end{eqnarray}
	For massless and electrically-neutral neutrinos, the contributions from diagrams (1), (2) or (4) are vanishing, since the relevant interaction vertices are proportional to either the fermion mass or the electric charge. 
	
	It is worthwhile to mention that although the obtained self-energies are seemingly different from those in Ref.~\cite{Denner:1991kt}, cf. Eqs.~(B.1)-(B.4) and (B.6)-(B.8) therein, they are actually identical after transforming the Passarino-Veltman functions ${\rm A}_0^{}$, ${\rm B}_{00}^{}$ and ${\rm B}_1^{}$ into ${\rm B}_{0}^{}$. With these self-energies, we can fix all the counterterms as in Eqs.~(\ref{eq:gauge_higgs_ct})-(\ref{eq:theta_w_ct}).
	
	\subsection{Amplitudes from the Counterterms}
	
	The counterterms result in new interaction vertices and additional diagrams to the scattering amplitudes of our interest. The Feynman rules for the counterterms have been derived in the previous literature~\cite{Denner:1991kt,Bohm:1986rj,Aoki:1982ed,Hollik:1988ii}, and the amplitudes from the counterterms can be easily obtained.
	
	\subsubsection{Self-energies of Gauge Bosons}
	
	The mass and wave-function counterterms of gauge bosons induce the following contribution
	\begin{eqnarray}
		{\rm i} \left(m_{Z,W}^2 \delta Z_{ZZ,W}^{} + \delta m_{Z,W}^2\right) g_{\mu\nu}^{}\;,
	\end{eqnarray}
	where $p^2 = 0$ has been assumed for the intermediate gauge bosons in the case of forward scattering. Furthermore, considering the external fermions, one obtains the scattering amplitudes of $\nu_\alpha^{} + f  \to \nu_\alpha^{} + f$  from the self-energy counterterms 
	\begin{eqnarray}
		{\rm i}{\cal M}_{{\rm c}}^{Z} &=& \frac{{\rm i}g^2}{4m_Z^4 c^2} \left(m_Z^2 \delta Z_{ZZ}^{} + \delta m_Z^2\right) \overline{\nu_\alpha^{}} \gamma_\mu P_{\rm L}^{} \nu_\alpha^{}\ \overline{f}\gamma^\mu\left(c_{\rm V,NC}^f-c_{\rm A,NC}^f\gamma^5\right)f  \;, \\
		{\rm i}{\cal M}_{{\rm c}}^{W} &=& \frac{{\rm i}g^2}{4m_W^4} \left(m_W^2 \delta Z_{W}^{} + \delta m_W^2\right) \overline{\nu_\alpha^{}} \gamma_\mu P_{\rm L}^{} \nu_\alpha^{}\ \overline{f}\gamma^\mu \left(c_{\rm V,CC}^f - c_{\rm A,CC}^f \gamma^5 \right) f \;.
	\end{eqnarray}
	
	\subsubsection{Vertex Counterterms}
	\begin{figure}[t]
		\centering
		\includegraphics[scale=1]{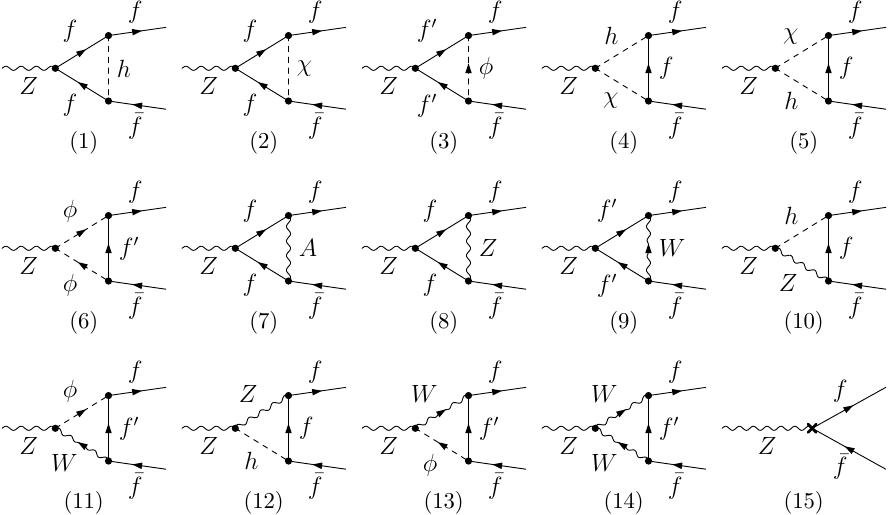}
		\caption{The one-loop diagrams for the corrections to the $f$-$f$-$Z$ vertex and the corresponding counterterm in the diagram (15).}
		\label{fig:ffZ}
	\end{figure}
	\begin{figure}[h]
		\centering
		\includegraphics[scale=1]{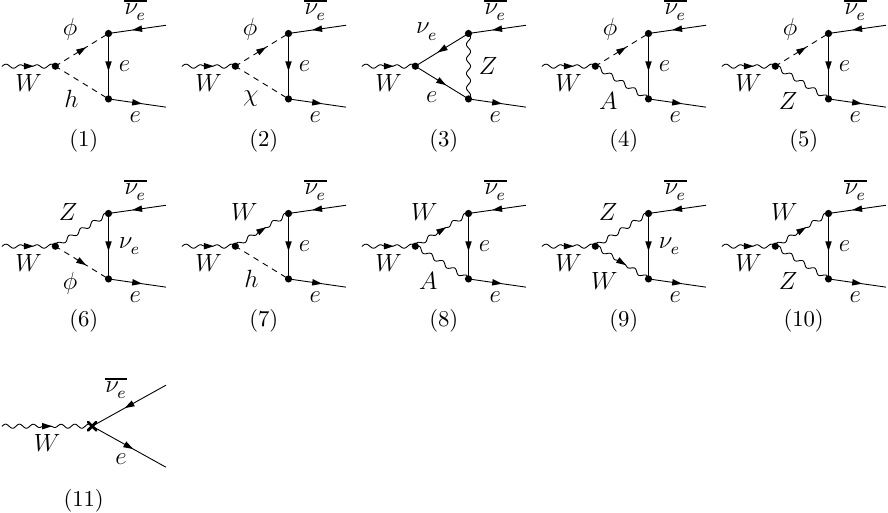}
		\caption{The one-loop diagrams for the corrections to the $\nu_e^{}$-$ e$-$ W$ and the corresponding counterterm in the diagram (11).}
		\label{fig:w_nue_e}
	\end{figure}
	
	The general fermion-vector-boson interaction from the counterterms can be expressed as
	\begin{eqnarray} \label{eq:FFV_ct}
		\delta \Gamma^{FFV}_\mu = {\rm i} e \gamma_\mu \left({\cal C}^-_f P_{\rm L}^{} + {\cal C}^+_f P_{\rm R}^{} \right)\;,
	\end{eqnarray}
	with $F$ stands for relevant fermions interacting with a given gauge boson $V$. All the one-loop diagrams for the corrections to the $f$-$f$-$Z$ vertex have been shown in Fig.~\ref{fig:ffZ}. The coefficients in front of chiral projection operators are defined as 
	\begin{eqnarray}
		{\cal C}^{\pm}_f = g_{f}^{\pm}\left(\frac{\delta g_{f}^{\pm}}{g_{f}^{\pm}}+\frac{1}{2} \delta Z_{Z Z}^{} + \delta Z_{i i}^{f, {\rm R(L)}} \right) + \frac{1}{2} Q_{f}^{} \delta Z_{A Z}^{} \;, 
	\end{eqnarray}
	where 
	\begin{eqnarray}
		g_{f}^{+} &=& -\frac{s}{c} Q_f^{}\;,\qquad \qquad \delta g_{f}^{+}=-\frac{s}{c} Q_f^{}\left(\delta Z_{e}+\frac{1}{c^{2}} \frac{\delta s}{s}\right)\;, \\ 
		g_{f}^{-} &=& \frac{I_{f}^{3} - s^{2} Q_f^{}}{sc}\;, \qquad \delta g_{f}^{-}=\frac{I_{f}^{3}}{s c}\left(\delta Z_{e}+\frac{s^{2}-c^{2}}{c^{2}} \frac{\delta s}{s}\right)+\delta g_{f}^{+} \;,
	\end{eqnarray}
	with the weak isospin generator $I_f^3$ of the SM fermions. The scattering amplitude from the counterterms turns out to be
	\begin{eqnarray}
		{\rm i}{\cal M}_{\rm c}^{\Gamma} = \frac{-{\rm i} g^2 s}{2m_Z^2 c} \left[\overline{\nu_\alpha^{}} \gamma_\mu {\cal C}^-_{\nu_\alpha^{}} P_{\rm L}^{} \nu_\alpha^{} \ \overline{f} \gamma^\mu \left(c_{\rm V,NC}^f - c_{\rm A,NC}^f \gamma^5\right) f + \overline{f} \gamma_\mu \left({\cal C}^-_f P_{\rm L}^{} + {\cal C}^+_f P_{\rm R}^{} \right) f\ \overline{\nu_\alpha^{}} \gamma^\mu P_{\rm L}^{} \nu_\alpha^{} \right] , \qquad
	\end{eqnarray}
	from which one can see that some corrections to the vector-type coupling are proportional to the tree-level coupling $c_{\rm V,NC}^f$ whereas others are not. 
	
	Several comments on Fig.~\ref{fig:ffZ} are helpful. For massless and electrically-neutral neutrinos, the contributions from the diagrams (1), (2), (4), (5), (7), (10) or (12) are vanishing. As the corrections of ${\cal O}(x_f^{})$ for $f=u,d,e$ are highly suppressed, the contributions from those diagrams can also be neglected. The flavor-dependent terms in the vertex correction come from the diagrams (3), (6), (9), (11), (13) and (14), which are consistent with the observations in Refs.~\cite{Botella:1986wy,Mirizzi:2009td}. Meanwhile, since neutrinos are purely left-handed in the SM, only ${\cal C}^-_{\nu_\alpha^{}}$ takes part in the correction. 
	
	The one-loop diagrams for the corrections to the $\nu_e^{}$-$ e $-$W$ vertex are given in Fig.~\ref{fig:w_nue_e}. The counterterm is similar to that in Eq.~(\ref{eq:FFV_ct}) but with 
	\begin{eqnarray}
		{\cal C}^{-}_f = \frac{1}{\sqrt{2}s}\left[\delta Z_e^{} - \frac{\delta s}{s} + \frac{1}{2} \delta Z_W^{} + \frac{1}{2}\left(\delta Z_{ii}^{\alpha,{\rm L}}+\delta Z_{ii}^{\nu_\alpha^{},{\rm L}}\right)\right] \;, \qquad {\cal C}^{+}_f = 0 \;. 
	\end{eqnarray}
	As the diagrams with the vertices proportional to the electron mass can be neglected, we just concentrate on those in (3), (8), (9) and (10).
	
	\subsection{Box Diagrams}
	\begin{figure}[t]
		\centering
		\includegraphics[scale=0.85]{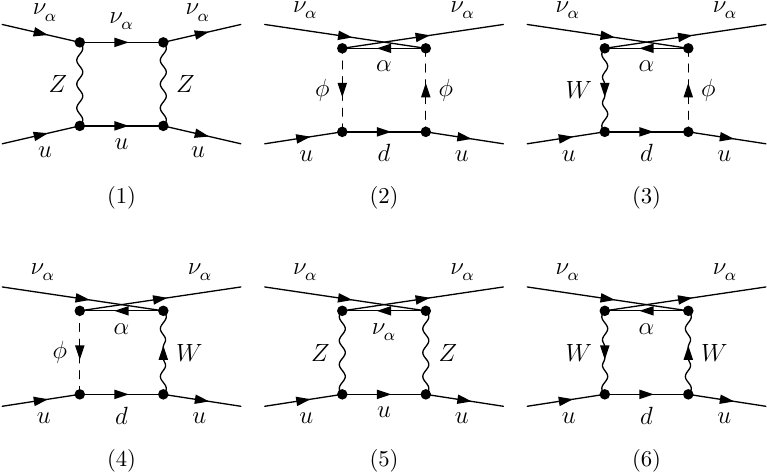}
		\caption{The one-loop box diagrams for the $\nu_\alpha^{} + u \to \nu^{}_\alpha + u$ scattering.}
		\label{fig:nu_u_box}
	\end{figure}
	
	\begin{figure}[h]
		\centering
		\includegraphics[scale=0.85]{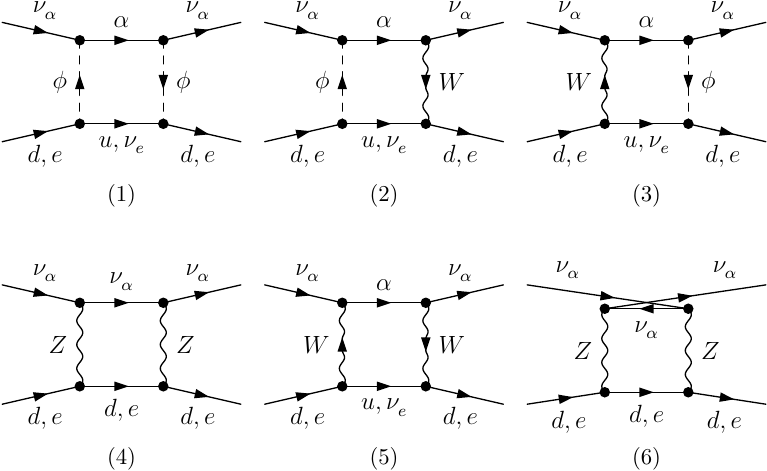}
		\caption{The one-loop box diagrams for the $\nu_\alpha^{} + d \to \nu^{}_\alpha + d$ and $\nu_\alpha^{} + e \to \nu^{}_\alpha + e$ scattering. For the electrons, only the NC contributions similar to quarks have been shown.}
		\label{fig:nu_d(e)_box}
	\end{figure}
	
	\begin{figure}[h]
		\centering
		\includegraphics[scale=1]{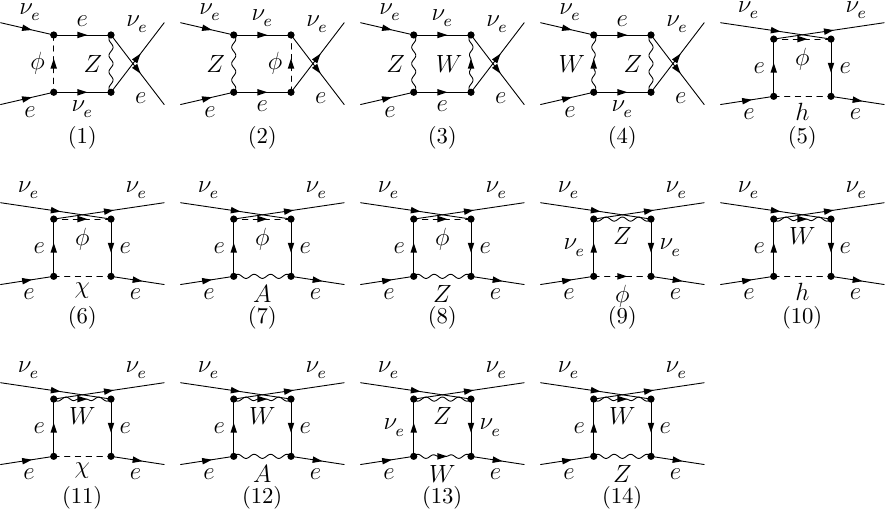}
		\caption{The one-loop box diagrams for the $\nu_e^{} + e \to \nu^{}_e + e$ scattering, where the contributions unique for electron neutrinos are retained and categorized as the CC type.}
		\label{fig:nu_e_box_CC}
	\end{figure}
	The box diagrams are presented in Figs.~\ref{fig:nu_u_box}, \ref{fig:nu_d(e)_box} and \ref{fig:nu_e_box_CC}, which are actually UV-finite. The final results of the amplitudes have been given and discussed in the main text. Notice that the diagrams involving $W$ or $\phi$ lead to the flavor-dependent corrections. 
	
	To simplify the expressions, one can expand the analytical formulas around the small fermion masses. However, there are two types of small fermion masses, namely, the charged-lepton masses and light quark masses. Given the strong mass hierarchy, i.e., $m_e^{} \ll m_u^{} \approx m_d^{}  \ll m_\mu^{} \ll m_\tau^{}$, we should first expand the results around $m^{}_{u,d}=0$ and $m^{}_e = 0$ and safely neglect ${\cal O}(x_{u,d,e})$ terms. 
	
	\bibliographystyle{elsarticle-num}
	\bibliography{ref}
	
\end{document}